\documentclass[notitlepage,twocolumn,pre,tightenlines,superscriptaddress,showpacs,floatfix]{revtex4-1} 

\usepackage{amsmath,amssymb,amsfonts,latexsym}
\usepackage{ulem}
\usepackage{bbold}
\usepackage{graphicx}
\usepackage{epsfig}
\usepackage{color}
\usepackage[activate=normal]{pdfcprot}  
\usepackage{bm} 
\usepackage{psfrag}
\usepackage{dsfont}
\usepackage[caption = false]{subfig}

\newcommand{\be}{\begin{equation}}
\newcommand{\ee}{\end{equation}}
\newcommand{\ben}{\begin{equation*}}
\newcommand{\een}{\end{equation*}}
\newcommand{\ba}{\begin{eqnarray}}
\newcommand{\ea}{\end{eqnarray}}

\newcommand{\nyuphysics}{Center for Soft Matter Research, Department of Physics, New York University, New York 10003, USA}
\newcommand{\nyuchemistry}{Department of Chemistry, New York University, New York 10003, USA}

\newcommand{\technionphy}{Department of Physics, Technion-IIT, 32000 Haifa, Israel}

\begin{document}

\title{Emergent Synchronization and Flocking in Purely Repulsive Self-Navigating Particles}

\author{Mathias Casiulis}
\email{mc9287@nyu.edu}
\affiliation{\technionphy}
\affiliation{\nyuchemistry}
\affiliation{\nyuphysics}
\author{Dov Levine}
\email{levine@technion.ac.il}
\affiliation{\technionphy}

\date{\today}

\begin{abstract}
Inspired by groups of animals and robots, we study the collective dynamics of large numbers of active particles, each one trying to get to its own randomly placed target, while avoiding collisions with each other.  The particles we study are repulsive homing active Brownian particles (HABPs) - self-propelled particles whose orientation relaxes at a finite rate towards an absorbing target in $2d$ continuous space.
For a wide range of parameters, these particles form synchronised system-wide chiral flocks, in spite of the absence of explicit alignment interactions.
We show that this dramatic behavior obtains for different system sizes and density, that it is robust against the addition of noise, polydispersity, and bounding walls, and that it can exhibit dynamical topological defects.
We develop an analogy to an off-lattice, ferromagnetic XY model, which allows us to interpret the different phases, as well as the topological defects.
\end{abstract}

\maketitle

\section{Introduction}

\begin{figure}[htbp]
    \centering
    \includegraphics[width = 0.92\columnwidth]{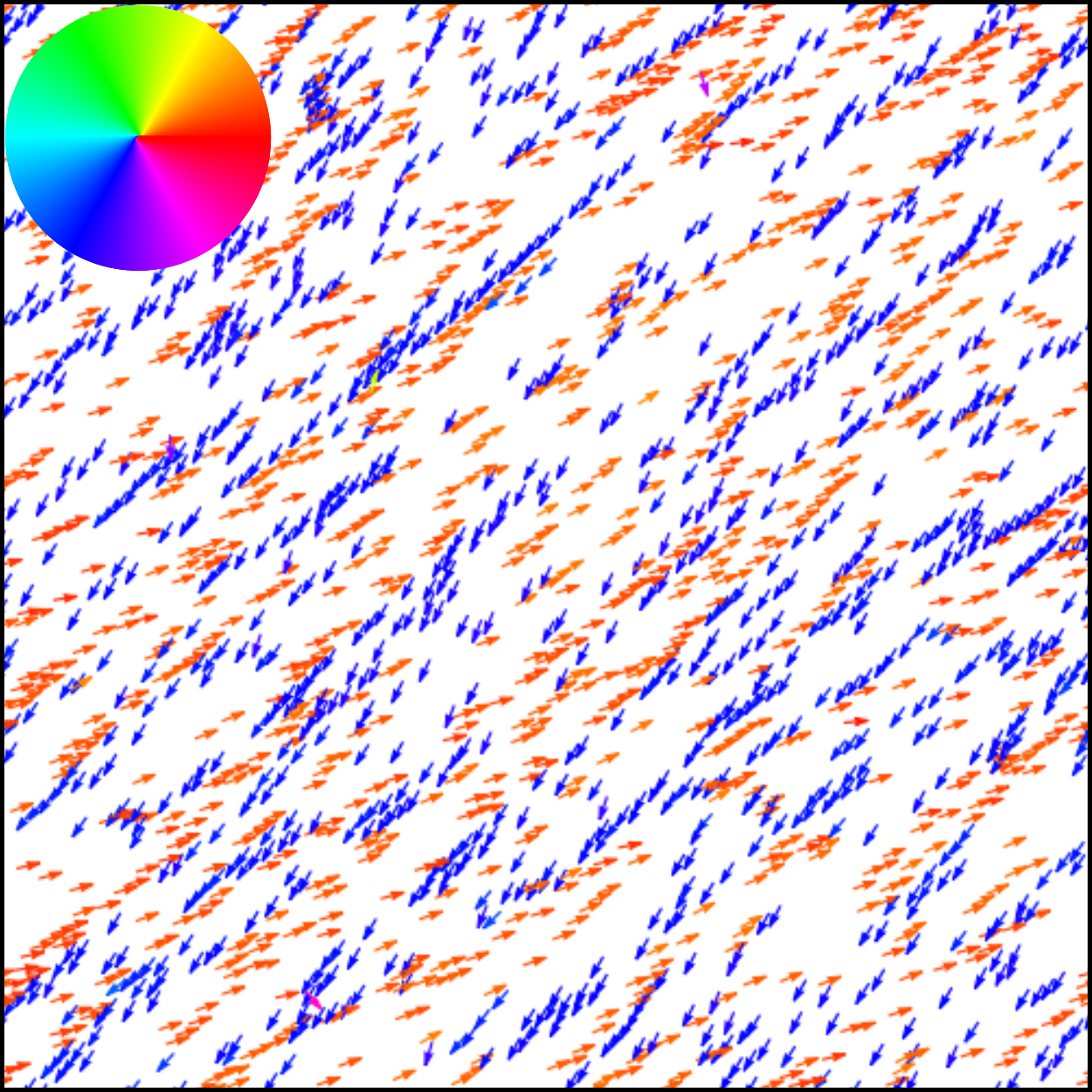}\\
    \vspace{0.1cm}
    \includegraphics[width = 0.92\columnwidth]{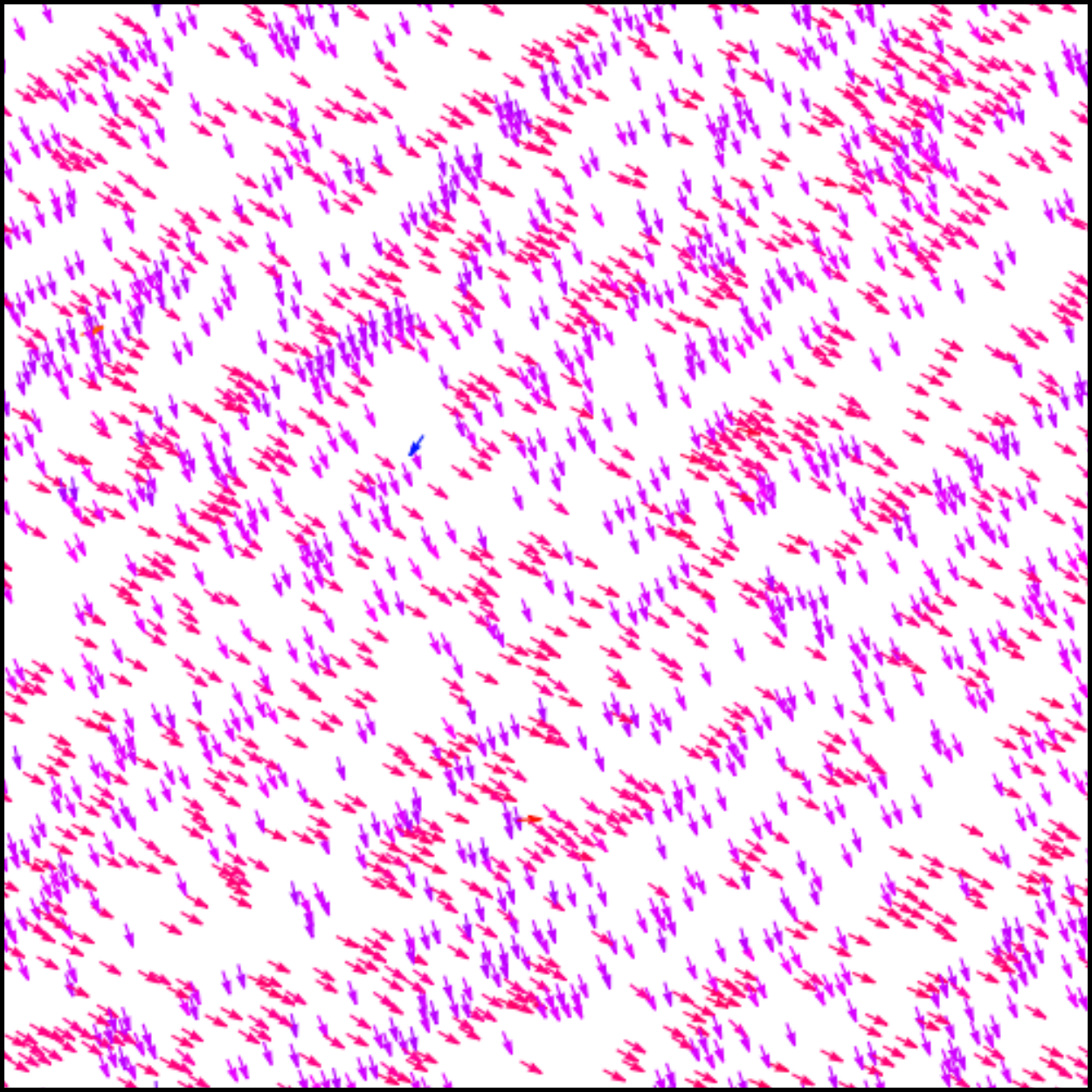}
    \caption{\textbf{Chiral synchronization.}
    Two snapshots, separated by a time $\Delta t = 50$, of a system of $N = 2048$ particles at a packing fraction $\phi = 0.20$ and for a relaxation rate $\Omega_r = 0.02$.
    Each particle is represented by its instantaneous self-propulsion vector, with color encoding orientation on a color wheel, shown as an inset.
    }
    \label{fig:SexyPic}
\end{figure}
Synchronized swirling is a spectacular phenomenon seen in systems of many scales, from molecular filaments~\cite{Schaller2010,Sumino2012} to groups of cells~\cite{Riedel2005,Chen2017}, to macroscopic animals~\cite{Moussaid2012,Vicsek2012,Calovi2014,Sugi2019}.
Designing minimal sets of instructions such that groups of artificial agents, like robotic swarms, can display harmonious motion, avoiding collisions, without central control or extensive communication is a current challenge in robotics~\cite{Mai2020, Talamali2021}, and in particular autonomous transportation~\cite{Ma2016}.
In abstract models, velocity alignment, or flocking~\cite{Vicsek1995,Toner95,Chate2020}, occurs at low densities only when there are explicit synchronizing interactions~\cite{Chen2017,Levis2017,Levis2019PRE,Levis2019PRR}.
However, recent studies have shown that chiral self-propelled particles interacting via repulsion only could feature striking self-organization properties, building up long-range properties like hyperuniformity~\cite{Lei2019,Zhang2022}.
In this paper, we study another example of such self-organization: we show that a dilute system of self-propelled particles~\cite{Marchetti2013} forms large chiral groups that flock and rotate in synchrony, despite only interacting through short-ranged repulsion~\cite{Casiulis2021} -- a phenomenon that is best appreciated by watching videos in the SI. 
Moreover, we show this phenomenon to be robust against a variety of possible disturbances.  This shows that large-scale flocking can be achieved by unbiased local interparticle interactions alone, an idea which may find application in robotics.

The system we study consists of many particles, each of which has a specific randomly placed target towards which it tends.
The flocks which develop are a striking example of emergent self-organization, and are surprising for several reasons.
First, it is remarkable that particles manage to harmoniously avoid each other and sustain regular trajectories at finite densities, with interleaved layers of particles reminiscent of the low-density BML model~\cite{Biham1992};
see Fig. \ref{fig:SexyPic} and videos in the SI.
Second, they achieve system-wide synchronization, even though the model \textit{contains no explicit aligning interactions}, which are required to observe large flocks in dilute chiral active matter~\cite{Levis2017,Levis2019PRE,Levis2019PRR,Lei2019,Fruchart2021}.
Third, even in the presence of effective alignment, one usually expects the long-range synchronization of $2d$ driven rotators in $2d$ space to be precluded by Mermin-Wagner-like arguments when their interactions are short-ranged and isotropic~\cite{Grinstein1993}.

\section{Model}

Our model consists of $N$ disks of diameter $a$, typically simulated in a 2D square box with side-length $L$ with periodic boundary conditions, although we will discuss hard boundaries later.
To each particle we associate a stationary target disk of diameter $a$.
Initially, the positions $\{\bm{r}_i(0), \bm{r}_{T,i}\}$ of  particles and targets are distributed randomly and uniformly in the box, and we give every particle a random initial orientation $\{\theta_i(0)\}$.
Each particle then follows the overdamped equations of motion
\begin{align}
    \dot{\bm{r}}_i &= v_0 \hat{\bm{e}}(\theta_i) + \sum\limits_{j\neq i} \bm{F}_{ji} + \sqrt{2 D_0}\bm{\eta}_i, \nonumber \\
    \dot{\theta_i} &= \omega_r \left(\theta_{i,T} - \theta_i \right) + \sqrt{2 D_r}\xi_i, \label{eq:HABPDynamics}
\end{align}
where $v_0$ is a self-propulsion speed, $\hat{\bm{e}}(\theta_i)$ is a unit vector making an angle $\theta_i$ with the $x$ axis, $\theta_{i,T}$ points towards the target of particle $i$, and $\bm{F}_{ji} = F_0 (r_{ij} - a) \hat{\bm{r}}_{ij} \mathds{1}( r_{ij} < a)$ is a harmonic repulsion term.
The orientation angles $\theta_i$ relax towards the targets at a finite rate $\omega_r$, and are understood modulo $2 \pi$, so that  $-\pi< \theta_{i,T} - \theta_i < \pi$.
This choice of a harmonic relaxation of angles, rather than a sinusoidal one, is motivated by the idea that robots or animals would likely have a monotonically increasing correction to their heading as it goes off target.
$\bm{\eta}_i$ and $\xi_i$ are two sets of unit-variance, zero-mean white noise that are delta-correlated in time and independent of one another. 
$D_0$ and $D_r$ are translational and rotational diffusion constants that can be tuned to adjust the noise levels in both equations.
These equations, together with the rule that whenever a particle and its target touch, they annihilate and are replaced by a new pair uniformly drawn in space, define a model we termed Homing Active Brownian Particles (HABP)~\cite{Casiulis2021}.
Note that the coupling between a particle and its target is here considered to be independent of range, which would be the case of isolated animals or robots travelling to known destinations.

Employing adimensional time and space units: $t \to v_0 t / a$ and $r \to r/a$,
the system is described by a set of six dimensionless parameters: the number of particles $N$, the packing fraction $\phi = N \pi a^2 / (4 L^2)$, the dimensionless hardness of particles $f_0 = F_0 / v_0$, the Péclet number $Pe \equiv v_0 a / D_0$, its rotational equivalent $Pe_r \equiv v_0 / (a D_r)$, and the dimensionless relaxation rate $\Omega_r \equiv \omega_r a / v_0$.
In the limit $\Omega_r \to 0$, one recovers a model of Active Brownian Particles (ABPs)~\cite{Fily2012}.
We fix $f_0 = 100$, ensuring that particles never overlap significantly.
This model was shown to undergo a jamming transition~\cite{Casiulis2021} for $\phi > \phi_J \approx 0.23$, but here we focus on the low-density phases of the model, $\phi < \phi_J$, where it is always expected to be a homogeneous fluid, and on relaxation rates $\Omega_r < \Omega_r^C$ small enough that the particles do not reach their targets ballistically.

Let us first consider a single particle-target pair.
For certain initial conditions, the equations of motion admit a constant-speed, circular-orbit solution around the target, with $\theta_T - \theta = \pm \pi/2$, and constant $\dot{\theta}$.
The radius of these orbit solutions is $R_0 = 2 v_0 / (\pi \omega_r) = 2 a / (\pi \Omega_r)$, and the sign of the angular speed defines a chirality for the trajectory.
Although this circular motion is reminiscent of so-called circular swimmers~\cite{Levis2017,Levis2019PRE,Levis2019PRR,Lei2019,Lowen2020}, there are two crucial differences: $(i)$ in our case, circular motion is only one possible solution, requiring special initial conditions {(see App.\ref{app:SingleParticle})}, and $(ii)$ HABPs have no intrinsic chirality.

\section{Collective synchronization}
To get a sense of the emergent synchronized states, let us begin by considering the model at finite relaxation rates with no noise.  First, we determine the chirality $\chi$ of each particle by measuring whether its target lies to its left ($\chi = +1$) or to its right ($\chi = -1$), when orienting the particle along its self-propulsion (For closed orbits, $\chi = +1$ is counterclockwise, and $\chi = -1$ is clockwise.).  
At each point in time, we split the particles into two chiral groups, with $n_L$ having $\chi = +1$ and $n_R$ having $\chi = -1$, and  measure the degree of alignment within each group: $\bm{\sigma}_L = n_L^{-1}\sum  \hat{\bm{e}}(\theta_i) \delta_{\chi,+1}$ and $\bm{\sigma}_R = n_R^{-1}\sum  \hat{\bm{e}}(\theta_i) \delta_{\chi,-1}$.
Since opposite-chirality groups cannot synchronise with each other, we quantify the extent of the synchronization by the parameter $\sigma \equiv (n_L |\bm{\sigma}_L| + n_R |\bm{\sigma}_R|)/N$.
Low values of $\sigma$ reflect low synchronization, while a value of unity indicates that the entire system is synchronized. 

\begin{figure}
    \centering
    \includegraphics[height = 0.48\columnwidth]{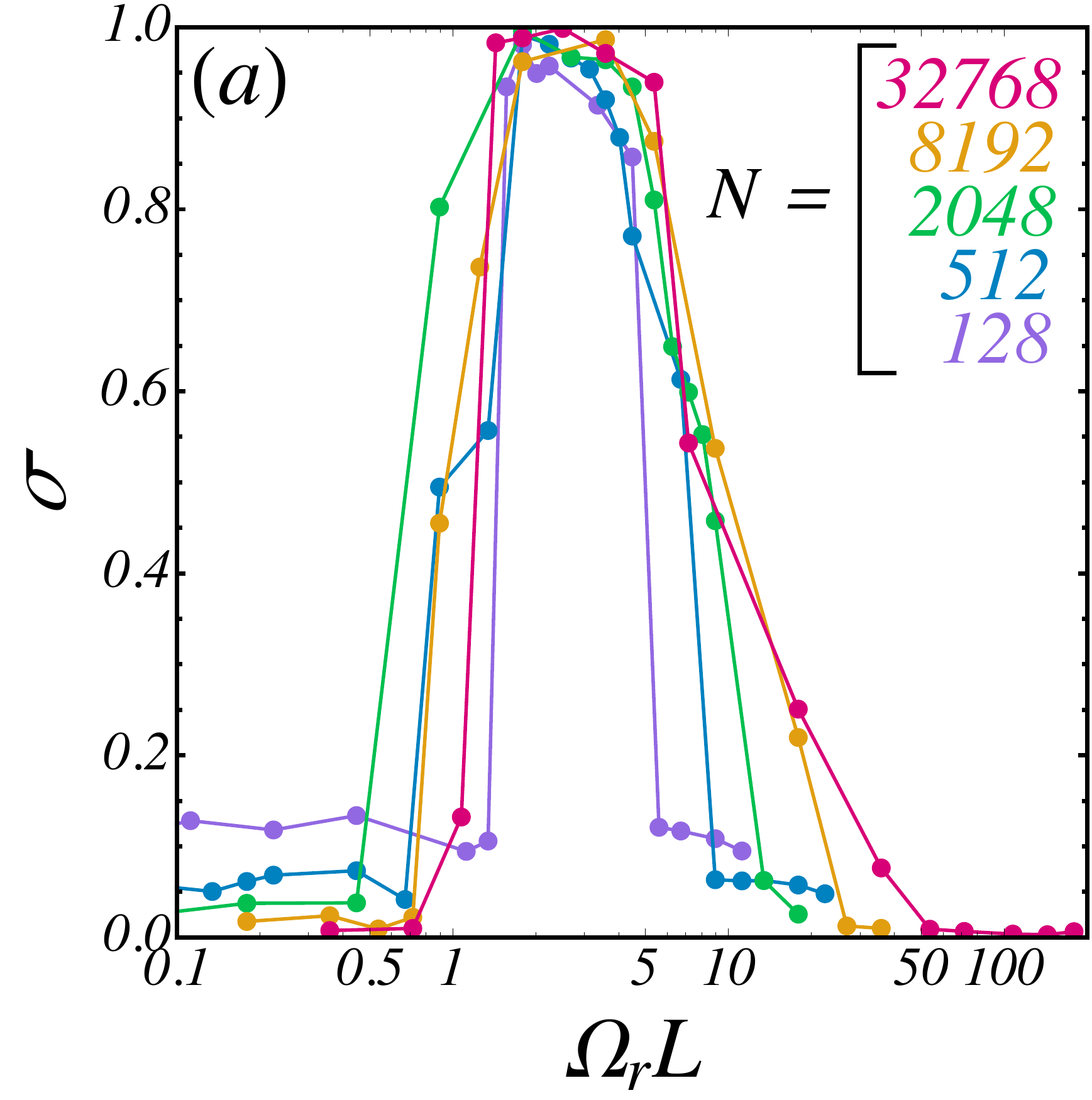}
    \includegraphics[height= 0.48\columnwidth]{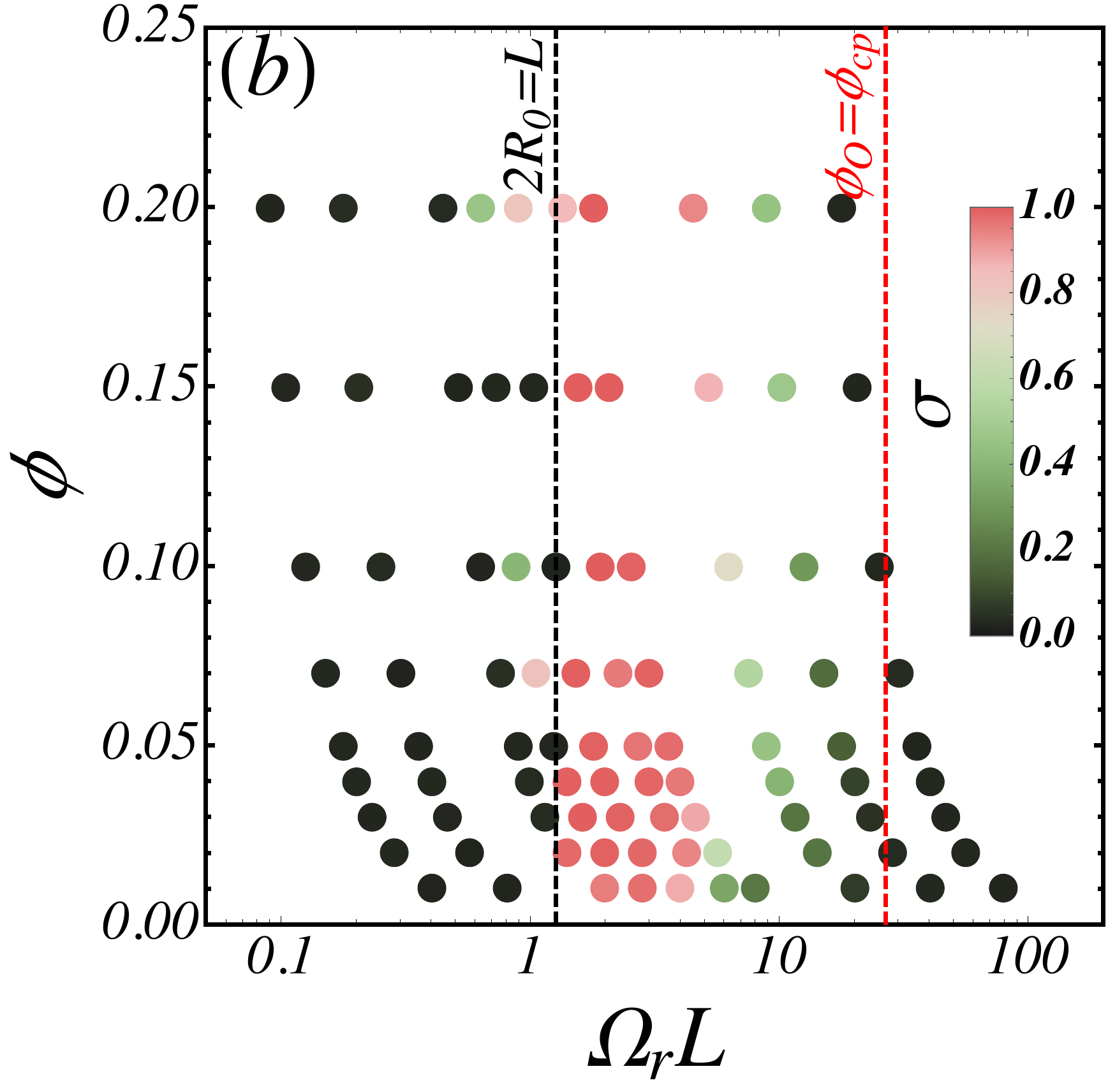}
    \caption{\textbf{Synchronization of noiseless HABPs.}
    $(a)$ Synchronization amplitude $\sigma$ against the rescaled relaxation rate $\Omega_r L$ at $\phi =0.2$ at several system sizes, averaged over 10 realisations.
    $(b)$ $2d$ map of the synchronization $\sigma$ in the $\Omega_r L$, $\phi$ plane for $N = 2048$. Lines represent $2 R_0 = L$ (black) and close-packing of orbits $\phi_O = \pi / 2 \sqrt{3}$.
    The color function used for $\sigma$ is shown in the color bar.
    }
    \label{fig:Noiseless}
\end{figure}

In Fig.~\ref{fig:Noiseless}$(a)$, we plot the steady-state value of $\sigma$ against $\Omega_r L$, when varying $\Omega_r$ at a fixed density $\phi = 0.2$ and for several system sizes $N$.
For small $\Omega_r$, $\sigma \approx 0$, with a sudden rise to $\sigma \approx 1$, indicating global synchronization, at $\Omega_r^{sync}\approx 1/L$.
$\sigma \approx 1$ is maintained for a while as $\Omega_r$ increases, eventually falling to zero again.
In this window the system is globally synchronized.

Some insight into this behavior comes by noting that orbits can only be stable in a bounded domain of $\Omega_r$ even at the single-particle scale.
On the one hand, since targets are absorbing, orbits can only persist if $R_0 > a$ (or, equivalently, $\Omega_r a < 2/\pi \approx 0.6$).
On the other hand, if the particle is placed in an $L\times L$ box, orbits must also satisfy $R_0 \lesssim L / 2$ ( $\Omega_r \gtrsim 4 a /\pi L \approx 1.3 a/L$).
Therefore, for a given system size, once can only have $\sigma > 0$ in an interval $4 a / (\pi L) \leq \Omega_r \leq 2 / \pi$ whose lower bound decreases as $L^{-1}$ and whose upper bound is roughly constant (see App.~\ref{app:RawCurves} for raw curves of $\sigma$ against $\Omega_r$).
In the data, we do observe that the maximally synchronized state is indeed observed at $\Omega_r L/a \approx 1$ across more than two orders of magnitude of $N$, see Fig.~\ref{fig:Noiseless}$(a)$, while the synchronization vanishes within an interval $\Omega_r \in \left[0.2;0.4\right]$ across the same range of sizes, slightly below the highest possible upper bound given above but still roughly constant.
This  holds at any density below $\phi_J$, as shown in Fig.~\ref{fig:Noiseless}$(b)$.
We note that this behavior is very different from that usually observed when tuning the strength of aligning interactions in phase oscillators~\cite{Strogatz2000}; that is, this model does not trivially map onto a Kuramoto model with $\Omega_r$ playing the role of the coupling.

At very low densities one could expect a sizeable domain of relaxation rates such that orbits are observed ($\Omega_r < 2 / \pi$) but do not interact ($\Omega_r L \gg 1$),
However, we find that synchronization decays far before orbits become trivially decoupled. 
This is seen in Fig.~\ref{fig:Noiseless}$(b)$, where we indicate with a red line the place where the packing fraction of orbits, $\phi_{O} = \phi 4 R_0^2/ a^2$, reaches the close-packing value $\phi_{cp} = \pi / \sqrt{12} \approx 0.91$.
The decay of synchronization happens at values of $\phi_O$ still large enough to ensure that the particles will interact, and only far to the right of this line could one observe absorbing states of independent orbits such as those described in previous works on circular swimmers~\cite{Lei2019}.

\section{Effect of noise}
\begin{figure}
    \centering
    \includegraphics[width = 0.48\columnwidth]{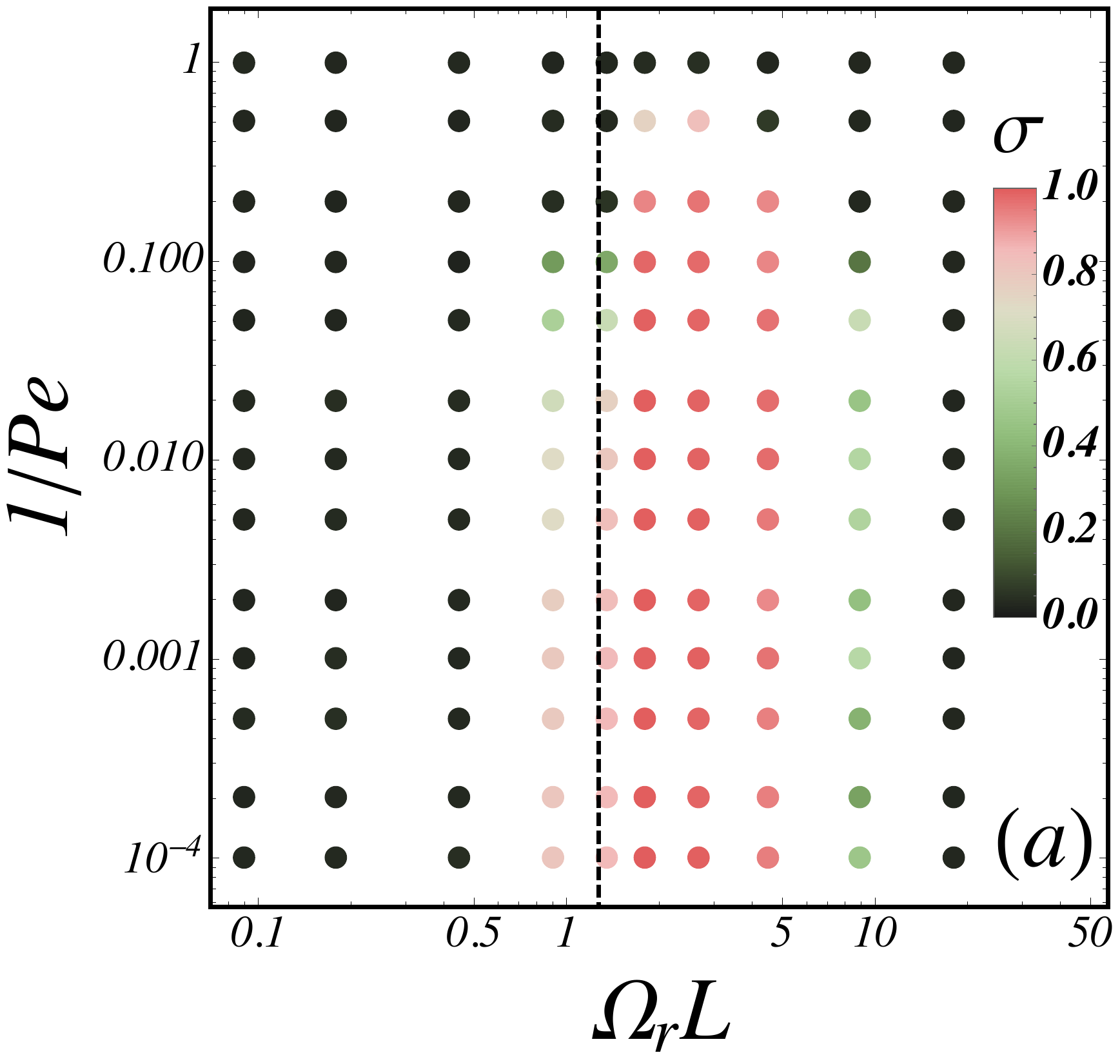}
    \includegraphics[width = 0.48\columnwidth]{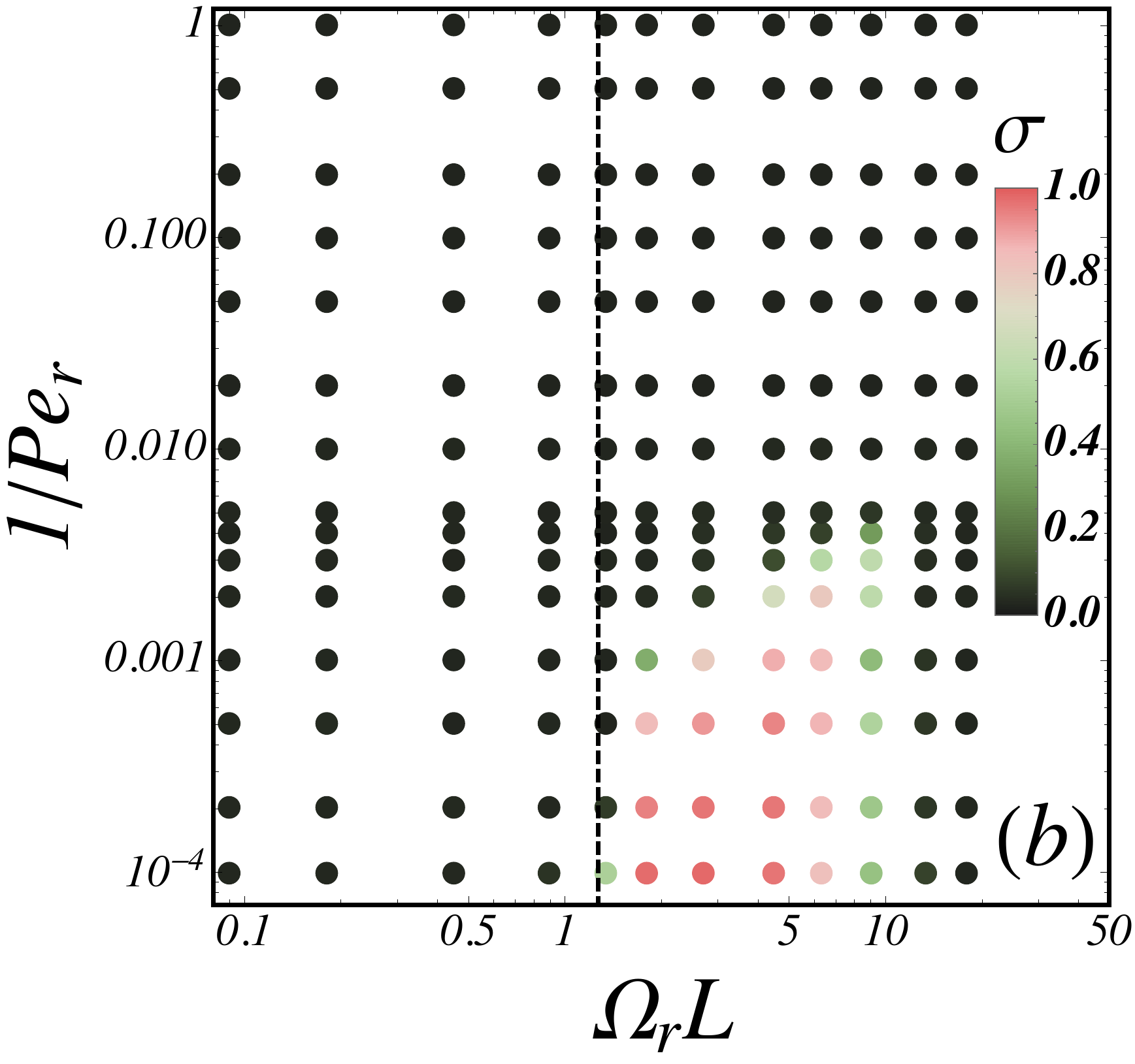}
    \caption{\textbf{Effects of noise}.
    $(a)$ Map of $\sigma$ in the rescaled relaxation rate $\Omega_r L$, dimensionless noise intensity $1/Pe$ plane, for $\phi = 0.2$ and $N = 2048$ particles. 
    $(b)$ Same map using rotational noise instead of translational.
    Vertical lines represent Lines represent $2 R_0 = L$.
    Here $\phi = 0.2$ and $N=2048$.
    }
    \label{fig:Noise}
\end{figure}

It is natural to inquire as to the effect of noise terms, as they are known to strongly affect traffic models like the BML model~\cite{Biham1992, Ding2011PRE, Ding2011JSM}.
Having verified (see App.~\ref{app:SingleParticle}) that a single isolated orbit survives a finite amount of noise, much like orbiting trajectories of confined active particles~\cite{Dauchot2019}, we show in Fig.~\ref{fig:Noise} that synchronization survives a finite amount of both translational and rotational noise.
For the case of translational noise, synchronization is destroyed at small $\Omega_r$ when diffusion displaces a particle of one orbit radius in one revolution; this happens when $D_0 \gg R_0^2\omega_r$, (or $1/Pe \gg 1/\Omega_r$).
In the case of rotational noise however, a smaller amount will destroy synchronization at lower relaxation amplitudes; this time, rotational diffusion needs to be directly compared to the relaxation rate, so that $D_r \gg \omega_r$ (or $1/Pe_r \gg \Omega_r$).
We also check (see App.~\ref{app:BinaryMixture}) that synchronization is remarkably robust against polydispersity in the relaxation rates, which can be thought of as some quenched behavioral noise.

\section{Effective model}
\begin{figure*}
    \centering
    \includegraphics[height = 0.48\columnwidth]{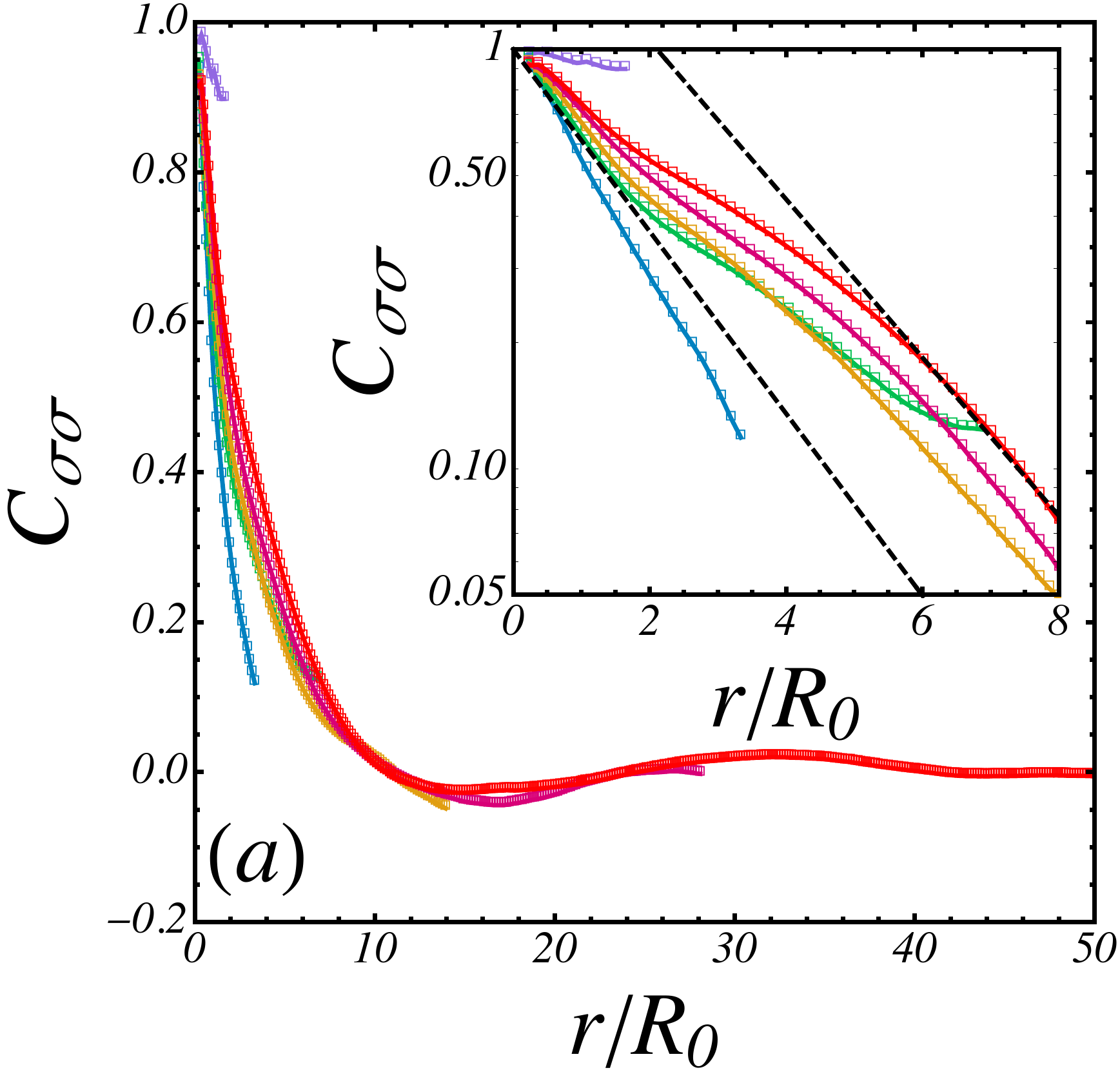}
    \includegraphics[height = 0.48\columnwidth]{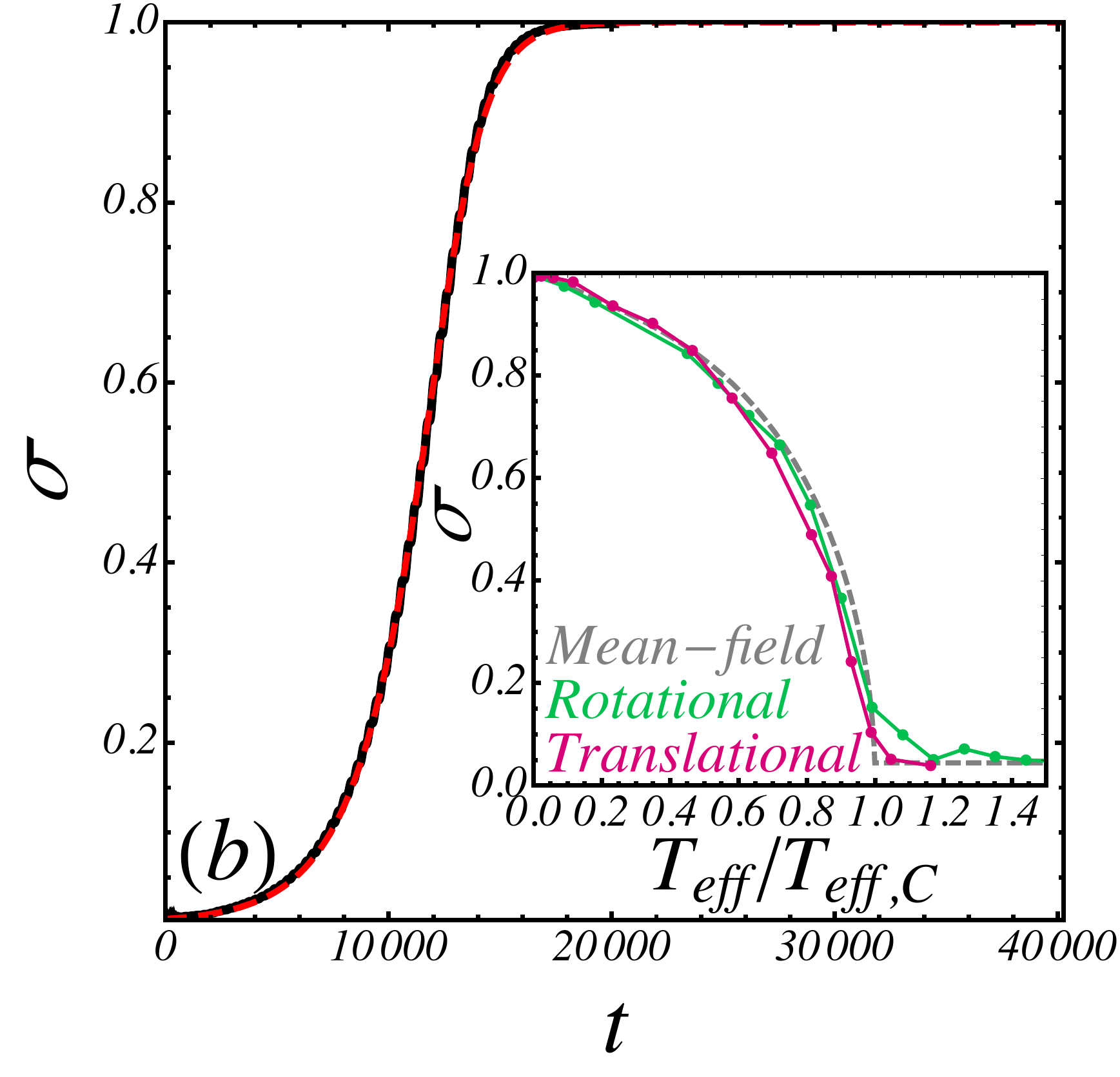}
    \includegraphics[height = 0.48\columnwidth]{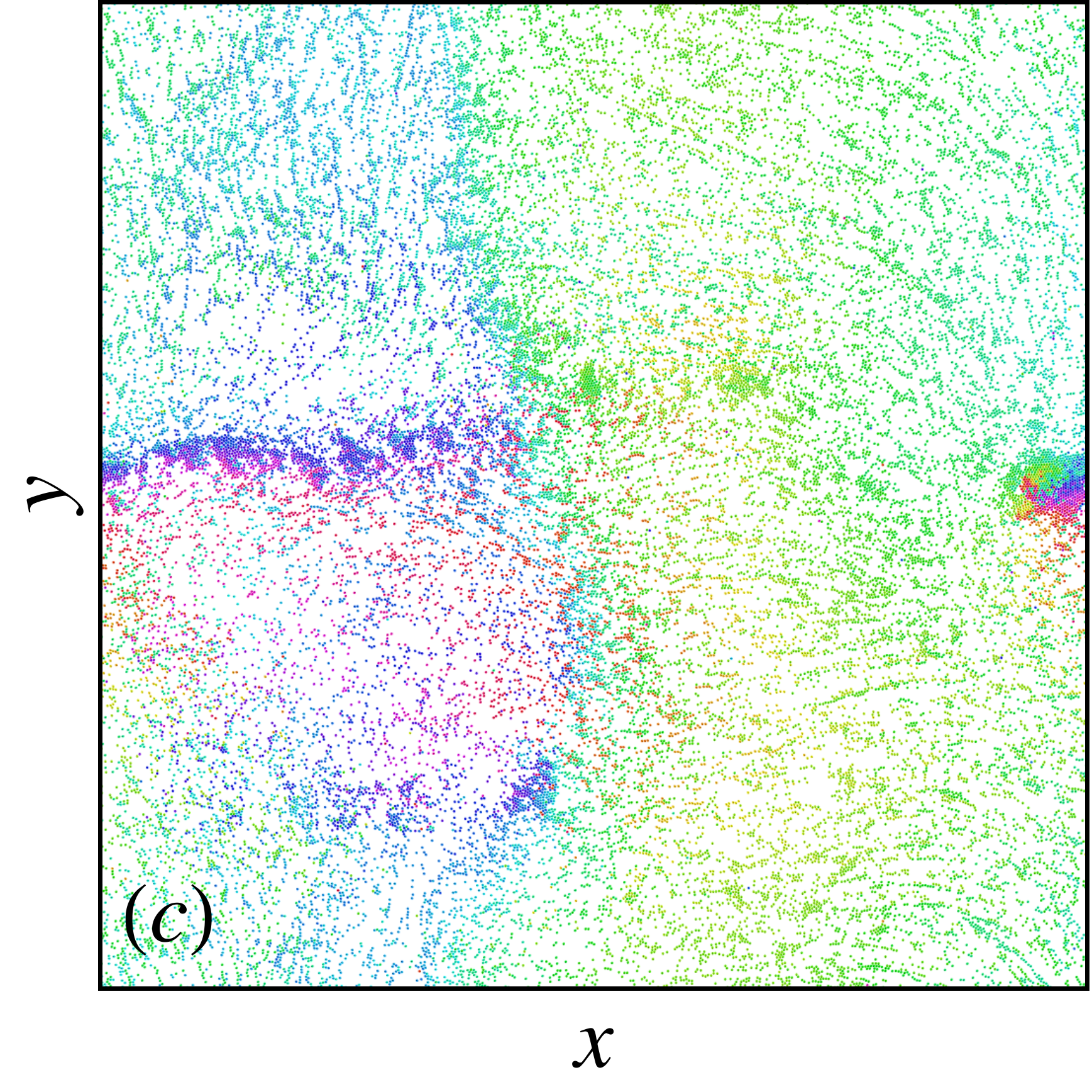}
    \includegraphics[height = 0.48\columnwidth]{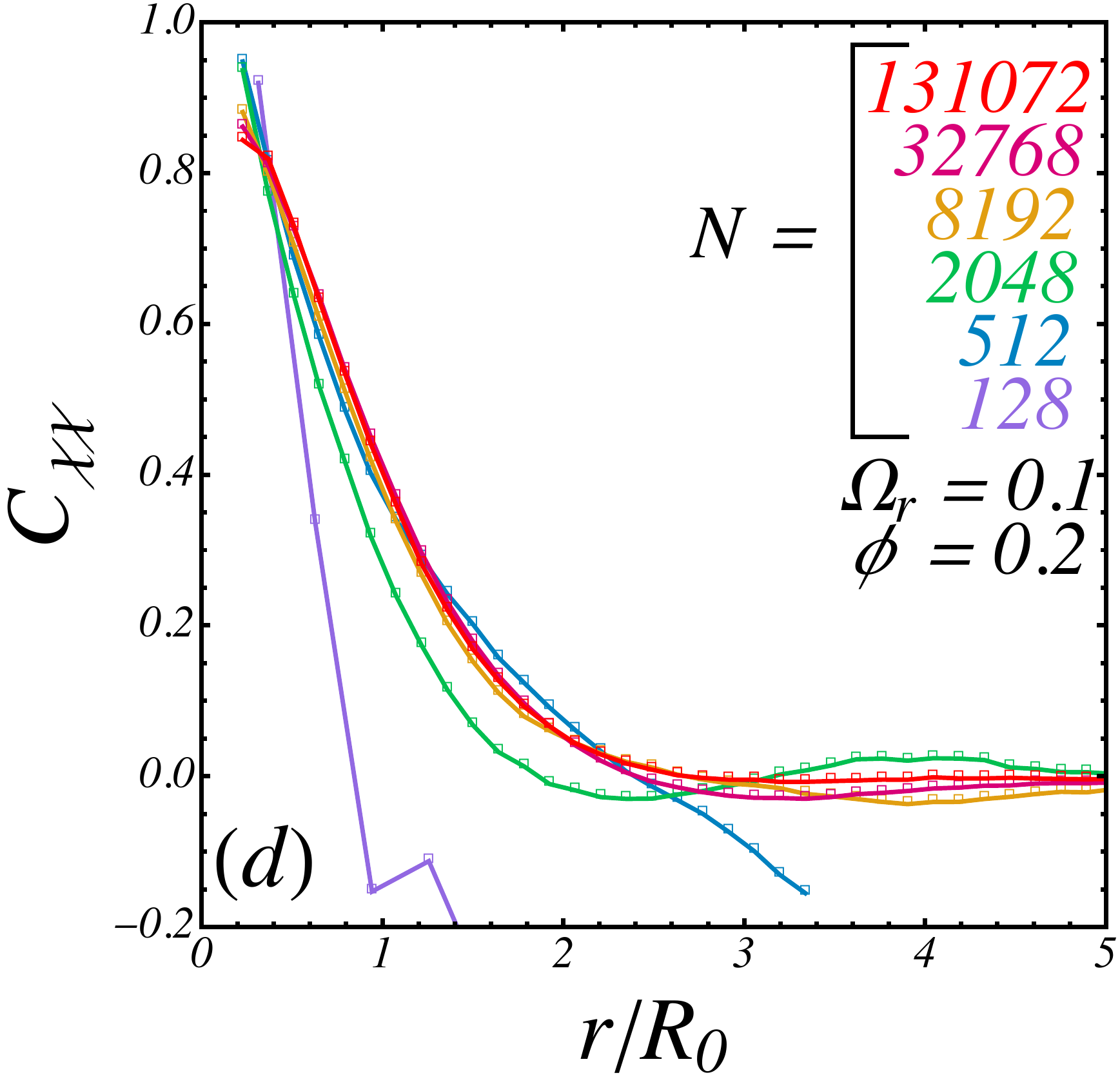}
    \caption{\textbf{Correlations.}
    $(a)$ Staggered correlation of self-propulsion orientations $C_{\sigma\sigma}$, against the distance $r$ over the orbit radius $R_0$, at different system sizes, averaged over $10$ to $100$ realisations.
    The color code of sizes is shown in $(d)$.
    They correspond to $\Omega_r L \approx 2.2, 4.5, 9.0, 18, 36, $ and $72$, respectively, from mauve to red.
    Inset: Zoom on the short-time decay, in log scale.
    Dashed lines show exponential decays with typical lengths $2 R_0$ and $2.5 R_0$.
    $(b)$ Transient dynamics of $\sigma$ starting from a uniform, random initial condition, for $N = 32768, \Omega_r = 0.007$ (black line) and best fit to the mean-field prediction (dashed red line).
    Inset: Steady-state $\sigma$ with rotational (green) or translational (red) noise, against the effective temperature $Pe_r^{-1}$ or $Pe^{-1}$, here noted $T_{eff}$, divided by its estimated critical value, for $N=2048$ particles at $\phi = 0.2, \Omega_r = 0.02$.
    The dashed gray line is the mean-field XY magnetization.
    $(c)$ Snapshot of a system showing topological defects, at $N=32768, \Omega_r = 0.02$. See full video in the SI.
    $(d)$ Correlation of chiralities, $C_{\chi \chi}$ against $r/R_0$ at different system sizes, using the same data as in $(a)$.
    }
    \label{fig:MeanField}
\end{figure*}
Having established the domain of existence of synchronization, we now seek to explain and characterise its buildup.
Self-propelled particles usually display velocity alignment or synchronization because they have explicit aligning interactions~\cite{Vicsek1995,Levis2017,Levis2019PRE,Levis2019PRR,Chate2020}.
Exceptions to this rule exist~\cite{NguyenThuLam2015}, but typically  at very high densities~\cite{Briand2018,Caprini2020,Szamel2021} or in confined geometries~\cite{Deseigne2012,Caprini2021b}.
In our model, we observe high polarisation of a continuous vector without any explicit alignment interaction, at rather low densities, and in full $2d$ space.

To understand this phenomenon, we first note that particles in our model are able to sustain stable circular orbits centered about their targets.
Once a particle-target pair reaches an orbit state, we may see it as analogous to a planar pendulum, moving in a circular orbit of fixed radius $R_0$ at a constant angular velocity $\pm \Omega_r$.
When driven pendula with similar orbits collide, they can synchronize provided that they have the same chirality and a small enough initial phase difference~\cite{Zhou2020}.
However, unless they interact via anisotropic interactions or at long range, driven phase oscillators are prevented from developing long-ranged alignment per a mapping~\cite{Grinstein1993} onto Kardar-Parisi-Zhang~\cite{Kardar1986} dynamics, that is effectively equivalent to a Mermin-Wagner~\cite{Mermin1966,Mermin1967,Mermin1968} argument.
It is therefore \textit{a priori} surprising to observe unit synchronization at wildly different values of $N$, as shown in Fig.~\ref{fig:Noiseless}$(a)$.

We can qualitatively describe this behavior with an effective coarse-grained model.
By symmetry, and considering only the leading orders in both synchronization amplitude and in the amplitude of spatial fluctuations, one would expect the dynamics of $\sigma$ to be captured by an effective coarse-grained free energy density $\mathcal{F}$ that only contains a $\varphi^4$ potential and a squared gradient~\cite{Grinstein1993,Kardar2007a},
\begin{align}
     \mathcal{F}[\sigma] \sim \frac{1}{2\tau}\left(\frac{\sigma^4}{4 \sigma_{\infty}^2} - \frac{\sigma^2}{2} + \frac{C}{2} (\bm{\nabla}\sigma)^2\right), \label{eq:FreeEnergy}
\end{align}
where $\tau$ is a characteristic time, and $\sigma_{\infty}$ and $C$ are functions of $\phi$ and $\Omega_r$.
In this effective theory, one expects $(\bm{\nabla}\sigma)^2 \sim \sigma^2 / \xi^2$, where $\xi$ is the correlation length of the synchronization.
Since the interaction range of an orbit is given by its radius, the correlation length $\xi$ should be proportional to $R_0$.
Thus, when $L/R_0 \sim 1$, as usual in finite systems smaller than their correlation length~\cite{Rulquin2016, Casiulis2019}, the gradient term becomes negligible and one is just left with a mean-field theory.
In other words, when the relaxation rate of HABPs is tuned, it affects the amplitude of the gradient term of the theory, sweeping all regimes from a mean-field theory to an XY model with short-ranged correlations.
Of course, in the limit of small correlation lengths, Eq.~\ref{eq:FreeEnergy} becomes less and less accurate, as other (more complicated) gradient terms reflecting the full microscopic couplings between orbits become relevant.

We confirm this picture in Fig.~\ref{fig:MeanField}.
First, we define the ``staggered'' correlation function of self-propulsion orientations within one chiral group,
\begin{align}
    C_{\sigma\sigma}(r) \equiv \frac{1}{c_0}\frac{\sum\limits_{i\neq j} (\hat{\bm{e}}(\theta_i) - \bm{\sigma}_{\chi_i})\cdot (\hat{\bm{e}}(\theta_j) - \bm{\sigma}_{\chi_j}) \delta_{\chi_i,\chi_j}\hat{\delta}(r - r_{ij})  }{\sum\limits_{i\neq j}\delta_{\chi_i,\chi_j}\hat{\delta}(r - r_{ij})},
\end{align}
where $c_0$ ensures that $C_{\sigma\sigma}(0) \to 1$,  $\delta_{\chi_i,\chi_j}$ is a Kronecker delta that selects same-chirality particles, $\bm{\sigma}(\chi_j)$ is the polarisation of the selected chirality, and $\hat{\delta}(r - r_{ij})$ is a binning function for the distances.
This function is similar in spirit to the velocity-velocity correlation functions defined in conventional flocks~\cite{Cavagna2010}.
This function is plotted in Fig.~\ref{fig:MeanField}$(a)$ at one relaxation rate ($\Omega_r = 0.1$) and density ($\phi = 0.2$) but several system sizes, against the distance in units of the radius of orbits, $R_0$.

In the case $N=128$, where the synchronization is very high, the correlation extends over the whole system, which mimics the long-range order predicted by mean-field theory.
This mean-field behavior can be checked by looking at the dynamics of the synchronization starting from random initial conditions.
Indeed, from (\ref{eq:FreeEnergy}) in the mean-field limit (no gradients), we expect 
\begin{align}
    \dot{\sigma}(t) &= \frac{1}{2\tau} \sigma(t) \left(1 - \frac{\sigma^2(t)}{\sigma_\infty^2} \right)
\end{align}
with solution
\begin{align}
    \sigma(t) &= \frac{\sigma_\infty}{\sqrt{1 + 3 e^{(t - t_0)/\tau}}},
\end{align}
which agrees well with the curves obtained in the high-synchronization regime; see Fig.~\ref{fig:MeanField}$(b)$.
We also check the behavior of $\sigma$ against rotational and translational noise amplitudes, each time rescaled by the estimated critical noise amplitude, choosing a relaxation amplitude such that $\sigma \approx 1$ in the noiseless case.
For both kinds of noise, the curves collapse and follow the mean-field magnetisation of an XY model~\cite{Kardar2007a}, confirming that the high-$\sigma$ regime displays mean-field-like behavior.

When the system gets larger, Fig.~\ref{fig:MeanField}$(a)$ shows a decay of the correlation, which eventually oscillates around zero at very large distances.
This decay occurs in two steps: a first decay occurs within one orbit diameter, and a second decay regime is observed beyond $2R_0$, showing that the scale for the synchronization decay is several orbit sizes.
Both decays scale exponentially with the distance, with a typical length of a few orbit diameters.

By analogy with the equilibrium $2d$ XY model~\cite{Berezinskii1971,Kosterlitz1973,Kosterlitz1974}, one would expect this exponential decay to be accompanied by the nucleation of topological defects in the system.
Such defects are indeed found in the regime $a \ll R_0 \leq L$, as shown in the snapshot of Fig.~\ref{fig:MeanField}$(c)$.
Note that these defects are observed at zero noise: varying the relaxation rate brings the system from a mean-field regime straight to a phase similar to that of high-temperature XY models with, seemingly, no critical phase in between.
During the dynamics (see videos in the SI), the centers of these defects play a special role, as they are associated with periodic accumulation of particles.

The correlation of chiralities can be studied by defining
\begin{align}
    C_{\chi\chi}(r) \equiv \frac{1}{c_{\chi,0}}\frac{\sum\limits_{i\neq j} (\chi_i - \overline{\chi}) (\chi_j - \overline{\chi}) \hat{\delta}(r - r_{ij})  }{\sum\limits_{i\neq j}\hat{\delta}(r - r_{ij})},
\end{align}
where $\overline{\chi}$ is the average chirality in the system.
In Fig.~\ref{fig:MeanField}$(d)$, we plot this and show that particles with the same chirality tend to stick together, but typically at a range shorter than $2R_0$. 
This correlation, as seen in snapshots of Fig.~\ref{fig:SexyPic} and videos of the dynamics (see SI), is actually also anisotropic, as particles tend to move in interleaved lanes of same-chirality particles.
These lanes are also responsible for the oscillations of $C_{\sigma\sigma}$ in panel Fig.~\ref{fig:MeanField}$(a)$.
While laning has been reported in traffic problems with driven entities heading in opposite~\cite{Tajima2002,Nagai2005,Moussaid2012,Poncet2017,Reichhardt2018} or perpendicular~\cite{Biham1992,Sun2018} directions, laning of circular trajectories is highly unusual.

\section{Closed Boundary Conditions}
\begin{figure}
    \centering
    \includegraphics[height=0.46\columnwidth]{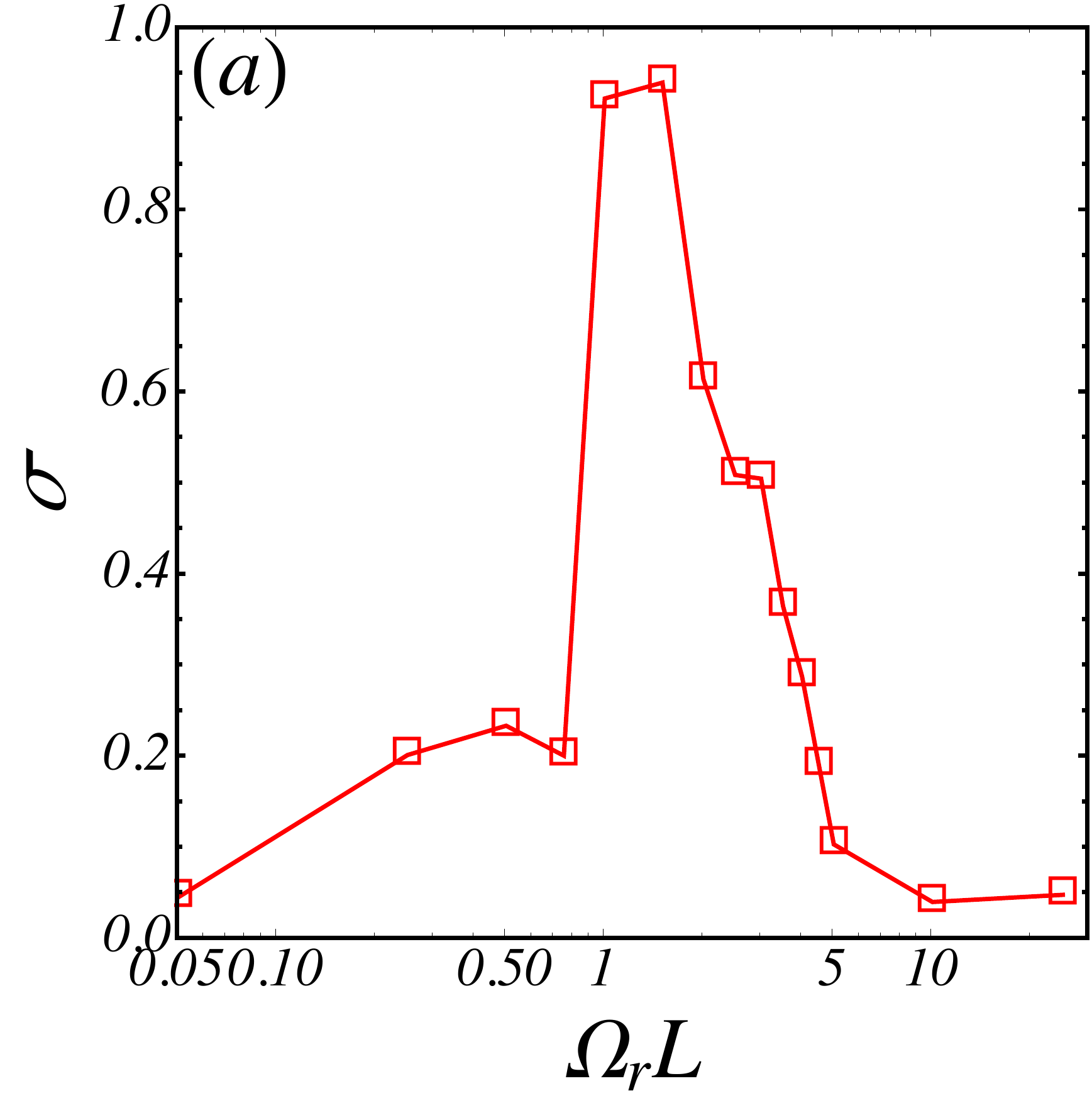}
    \includegraphics[height=0.46\columnwidth]{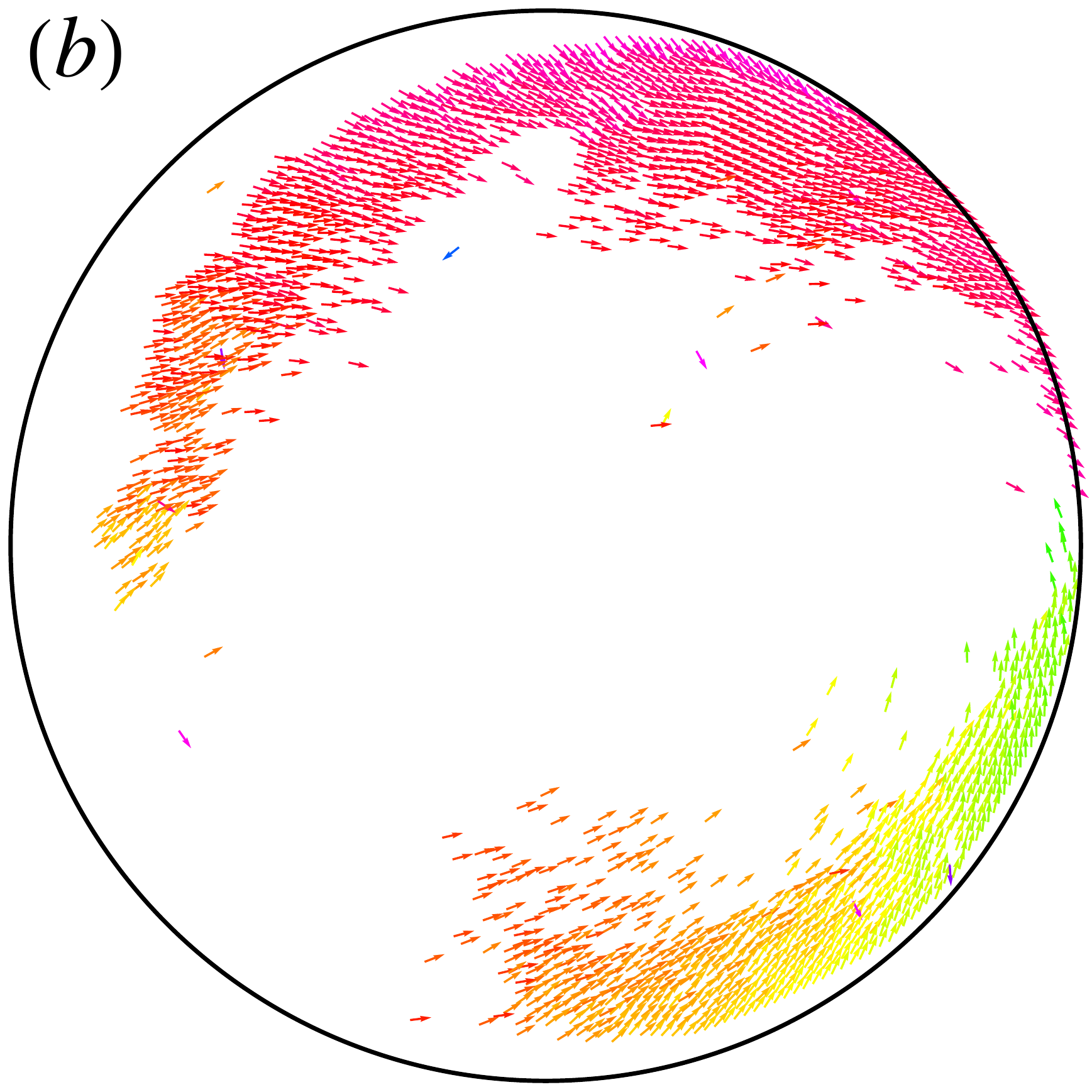}
    \caption{\textbf{Effect of walls.}
    $(a)$ Synchronization $\sigma$ against $\Omega_{r} L$ for a system of $N=2048$ particles at $\phi = 0.2$, when bounded by a hard circular wall with radius $L$.
    $(b)$ Snapshot of the system at the relaxation rate $\Omega_r = 0.02$ ($\Omega_r L \approx 1$).
    Each particle is represented by its instantaneous self-propulsion vector, with color encoding orientation.
    }
    \label{fig:BoundingWall}
\end{figure}

Finally, the addition of bounding walls can have dramatic effects on the build-up of density and velocity correlations in systems of self-propelled particles~\cite{Deseigne2010,Deseigne2012,BenDor2021, Codina2021}.
Therefore, we briefly check that the synchronization of HABPs survives when they are placed inside of a simulation box bounded by a hard, circular wall with radius $L$.
The results are shown in Fig.~\ref{fig:BoundingWall}.
In panel $(a)$, we show the synchronization amplitude $\sigma$ against the rescaled relaxation rate $\Omega_r L$ for $\phi = 0.2$ and $N=2048$.
It is essentially the same as in periodic boundary conditions.
Panel $(b)$ shows a snapshot of a well-synchronized configuration, which suggests that walls lead to self-sorting into swirling chiral flocks in lieu of simple laning, a fact that we also checked in a closed square with hard walls (see videos in the SI).

\section{Conclusions}
We have studied a minimal model of repulsive homing particles at low densities.
In spite of collisions, these particles eventually achieve stable circular orbits and manage to harmoniously avoid each other -- even in the presence of finite noise, polydispersity in behaviors, or bounding walls.
Despite the lack of any explicit aligning interactions, these orbits synchronize at the scale of their diameter, in a way reminiscent of the collective actuation of particles in active elastic solids due to rotations around their resting positions~\cite{Baconnier2021}.
Since particles and targets are coupled in a non-reciprocal way, this synchronization might be a distant relative of chiral synchronization of non-reciprocal rotators~\cite{Fruchart2021}.
Such remarkable self-organization emerging from simple ingredients suggest that harmonious collective motion is achievable with minimal communication in real systems of self-navigating agents, like animals or robots, without resorting to explicit alignment.

\acknowledgements{We thank Daniel Hexner, Yariv Kafri, and Olivier Dauchot for useful and insightful discussions.  D.~L. and M.~C. were supported by the Israel Science Foundation under grant No. 1866/16. M.~C. acknowledges funding from the Simons Foundation through the Simons Center for Computational Physical Chemistry, Department of Chemistry, NYU.}

\bibliography{PostDoc-DovLevine}

\appendix
\onecolumngrid

\section{Captions of the videos}

We here detail the content of the various ancillary video files.
All videos are animated at 30 frames per second, with a time between two frames $\delta t = 0.5 a / v_0$, where $a$ the repulsive diameter of particles and $v_0$ the self-propulsion speed.

\textit{FullBuildup.mp4 --} Short-time dynamics of a system of noiseless Homing Active Brownian particles, or HABPs, starting from a uniform local density of particles in space, and a uniform distribution of self-propulsion orientations, until a synchronised steady state builds up.
The number of particles is $N = 128$, the overall packing fraction $\phi = 0.2$, and the dimensionless relaxation rate $\Omega_r = 0.1$.
The colour of each particle codes for the phase of the instantaneous orientation of the self-propulsion (mapped onto a colour wheel like in the main text), which is also represented by a black arrow inside of each particle.

\textit{2k\_0p02\_highmag.mp4 --} Steady-state dynamics of a synchronised state of HABPs with a synchronization amplitude $\sigma \approx 1$. 
In this simulation, the number of particles is $N = 2048$, the density $\phi = 0.2$, the noise amplitudes are all set to zero, and the relaxation rate is $\Omega_r = 0.02$.
The colour code is the same as in the previous video.
Fig.~1 of the main text was made from snapshots of this video.

\textit{2k\_0p02\_LaneFlock\_BackseatView.mp4 --} Same data as in the previous video, but this time $3d$-rendered using a ray-tracing algorithm.
The video follows the perspective of a tagged particle, shown as a spherical light emitter in the middle of the frame, travelling its orbit in the midst of other particles, represented as colored refractive spheres with refraction index $n = 1.5$.
The colour code is the same as in the previous video, and the camera angle is chosen so that its azimuthal component always matches the $2d$ orientation of the self-propulsion of the tagged particle.

\textit{32k\_0p02\_intermediarymag.mp4 --} Steady-state dynamics of a synchronised state of HABPs, with an intermediary synchronization amplitude, $\sigma \approx 0.5$. 
In this simulation, the number of particles is $N = 32768$, the density $\phi = 0.2$, the noise amplitudes are all set to zero, and the relaxation rate is $\Omega_r = 0.02$.
The colour code is the same as in the previous video.

\textit{32k\_0p1\_lowmagnetisation.mp4 --} Steady-state dynamics of a synchronised state of HABPs, with a low synchronization amplitude, $\sigma \approx 0.1$. 
In this simulation, the number of particles is $N = 32768$, the density $\phi = 0.2$, the noise amplitudes are all set to zero, and the relaxation rate is $\Omega_r = 0.1$.
The colour code is the same as in the previous video.

\textit{2k\_bidisperse\_onlyone\_even.mp4 --} Steady-state dynamics of a system of HABPs with bidisperse relaxation rates. Here, each particle is represented by its instantaneous self-propulsion polarity, which is coloured according to two criteria: the gray arrows are particles with one of the relaxation rates, while the coloured arrows are the group with the other relaxation rate. The precise colour then represents the chirality of each particle: right-goers are green and left-goers are red.
In this simulation, the number of particles is $N = 2048$, the density $\phi = 0.2$, the noise amplitudes are all set to zero, and the relaxation rates are $\Omega_{r,1} = 0.02$ for the coloured arrows (very synchronized on large orbits), and $\Omega_{r,2} = 0.01$ for gray arrows (not on orbits).

\textit{2k\_bidisperse\_onlyone\_odd.mp4 --} Same video as above, but this time colouring the non-synchronized population: left-goers are blue and right-goers are orange.

\textit{2k\_bidisperse\_both\_even.mp4 --} Same as \textit{2k\_bidisperse\_onlyone\_even.mp4}, but with both families of particles undergoing synchronized motion, at two different frequencies.
In this simulation, the number of particles is $N = 2048$, the density $\phi = 0.2$, the noise amplitudes are all set to zero, and the relaxation rates are $\Omega_{r,1} = 0.02$ for the coloured arrows (very synchronized on large orbits), and $\Omega_{r,2} = 0.05$ for gray arrows (very synchronized on smaller orbits).

\textit{2k\_bidisperse\_both\_odd.mp4 --} Same video as above, but this time colouring the small-orbit population: left-goers are blue and right-goers are orange.

\textit{2k\_bidisperse\_oneabsorbed\_even.mp4 --} Same as \textit{2k\_bidisperse\_onlyone\_even.mp4}, but with one fast-relaxing family of particles that reaches its targets in almost direct paths.
In this simulation, the number of particles is $N = 2048$, the density $\phi = 0.2$, the noise amplitudes are all set to zero, and the relaxation rates are $\Omega_{r,1} = 0.02$ for the coloured arrows (very synchronized on large orbits), and $\Omega_{r,2} = 0.2$ for gray arrows (not on orbits).

\textit{2k\_bidisperse\_oneabsorbed\_odd.mp4 --} Same video as above, but this time colouring the non-synchronized population: left-goers are blue and right-goers are orange.

\textit{2k\_phi0p2\_0p02\_walledchiralflocks.mp4 --} Steady-state dynamics of a synchronised state of HABPs with a synchronization amplitude $\sigma \approx 1$, placed within a circular hard wall (black). 
In this simulation, the number of particles is $N = 2048$, the density $\phi = 0.2$, the noise amplitudes are all set to zero, and the relaxation rate is $\Omega_r = 0.02$.
Colour codes for the orientation of self-propulsion
Fig.~5$(b)$ of the main text was made from snapshots of this video.

\textit{2k\_phi0p1\_0p015\_walledswirlingflocks.mp4 --} Steady-state dynamics of a synchronised state of HABPs with a synchronization amplitude $\sigma \approx 0.8$, placed within a circular hard wall (black). 
In this simulation, the number of particles is $N = 2048$, the density $\phi = 0.1$, the noise amplitudes are all set to zero, and the relaxation rate is $\Omega_r = 0.015$.
The colour code is the same as in the previous video.

\textit{2k\_phi0p2\_re0p02\_squarewalls.mp4 --} Steady-state dynamics of a synchronised state of HABPs with a synchronization amplitude $\sigma \approx 0.9$, placed within a square-shaped hard wall (black). 
In this simulation, the number of particles is $N = 2048$, the density $\phi = 0.2$, the noise amplitudes are all set to zero, and the relaxation rate is $\Omega_r = 0.02$.
The colour code is the same as in the previous video.

\textit{2k\_phi0p2\_re0p02\_InvPe0p2\_highmag.mp4 --} Steady-state dynamics of a synchronised state of HABPs with a synchronization amplitude $\sigma \approx 0.8$, placed within a square-shaped periodic box, with non-zero translational noise. 
In this simulation, the number of particles is $N = 2048$, the density $\phi = 0.2$, the translational noise is switched on at $1/Pe = 0.2$, and the relaxation rate is $\Omega_r = 0.02$.
The colour code is the same as in the previous video.

\textit{2k\_phi0p2\_re0p02\_InvPe0p0005\_highmag.mp4 --} Steady-state dynamics of a synchronised state of HABPs with a synchronization amplitude $\sigma \approx 0.8$, placed within a square-shaped periodic box, with non-zero rotational noise. 
In this simulation, the number of particles is $N = 2048$, the density $\phi = 0.2$, the rotational noise is switched on at $1/Pe_r = 0.0005$, and the relaxation rate is $\Omega_r = 0.02$.
The colour code is the same as in the previous video.

\section{Numerical methods\label{app:NumericalMethods}}

All the results presented in the main text are obtained via molecular dynamics (MD) simulations with the simplest possible order-1 integrator.
Namely, we write the equation of motion of any Cartesian component of the position of a particle symbolically as
\begin{align}
    dx = x(t+dt) - x(t) = v_{det} dt + v_{stoch}dt^{1/2},
\end{align}
where $dt$ is a fixed time step, $v_{det}$ is the deterministic part of the velocity that comes from self-propulsion and interactions with other particles, and $v_{stoch}$ is the stochastic part of the velocity that appears when we introduce noise.
In the case with noise, the stochastic part of the velocity simply reads $v_{stoch} = \sqrt{2 D_T} \eta_x$, with $\eta_x$ drawn from a unit-variance centered normal distribution, and it is zero otherwise.
The computation of the interaction part of $v_{det}$ is accelerated by introducing a partition of space into square cells twice as wide as the longest-range interaction in the system, and labelling at all times each particle with its cell number.
In practice, we set the time step to, at most, $dt = 10^{-4}$, or to the largest power of ten that ensures that no update $dx$ can be larger than $0.01$ in simulation units.
This choice ensures that even high noise amplitudes cannot simply bypass repulsive interactions, e.g. jump to the other side of a neighbouring particle, due to the choice of discretisation of time.
For instance, if $\sqrt{2 D_T} = 100$, corresponding in the main text to an inverse Péclet number $50$, we set $dt = 10^{-6}$.

The initial positions of particles and targets are each drawn uniformly in a periodic square simulation box with linear size $L$, only rejecting pairs such that particles are absorbed at drawing time.
When the relaxation rate of the self-propulsion orientation towards the target is finite, we also draw the initial polarity of each particle uniformly on the circle.

When hard walls are considered, we reflect any update leading outside of the box towards the inside of the box.
More concretely, for a particle starting at position $\bm{r}_0$ and a proposed move to $\bm{r}_0 + \bm{\delta r}$ lying outside the simulation box, we first move the particle to the point of contact with the wall, $\bm{r}_0 + \bm{\delta r}_c$, then compute the line tangent to both the particle and the wall at that point, and flip the remainder $\bm{\delta r}_r = \bm{\delta r} - \bm{\delta r}_c$ of the proposed displacement perpendicular to that line.
In other words, we operate the replacement:
\begin{align}
    \bm{\delta r}_r = \bm{\delta r}_r^{//} + \bm{\delta r}_r^{\perp} \to \bm{\delta r}_r^{\text{flip}} = \bm{\delta r}_r^{//} - \bm{\delta r}_r^{\perp},
\end{align}
where the exponents indicate the parallel and orthogonal directions with respect to the wall at contact.
We then update the position to $\bm{r}_0 + \bm{\delta r}_c + \bm{\delta r}_r^{\text{flip}} $.
If that last position lies outside of the box again, which is increasingly unlikely as $dt$ becomes smaller and as long as the walls are regular enough, we repeat the reflection steps.

\section{Synchronization amplitude against non-rescaled relaxation rate\label{app:RawCurves}}

\begin{figure}
    \centering
    \includegraphics[width=0.50\textwidth]{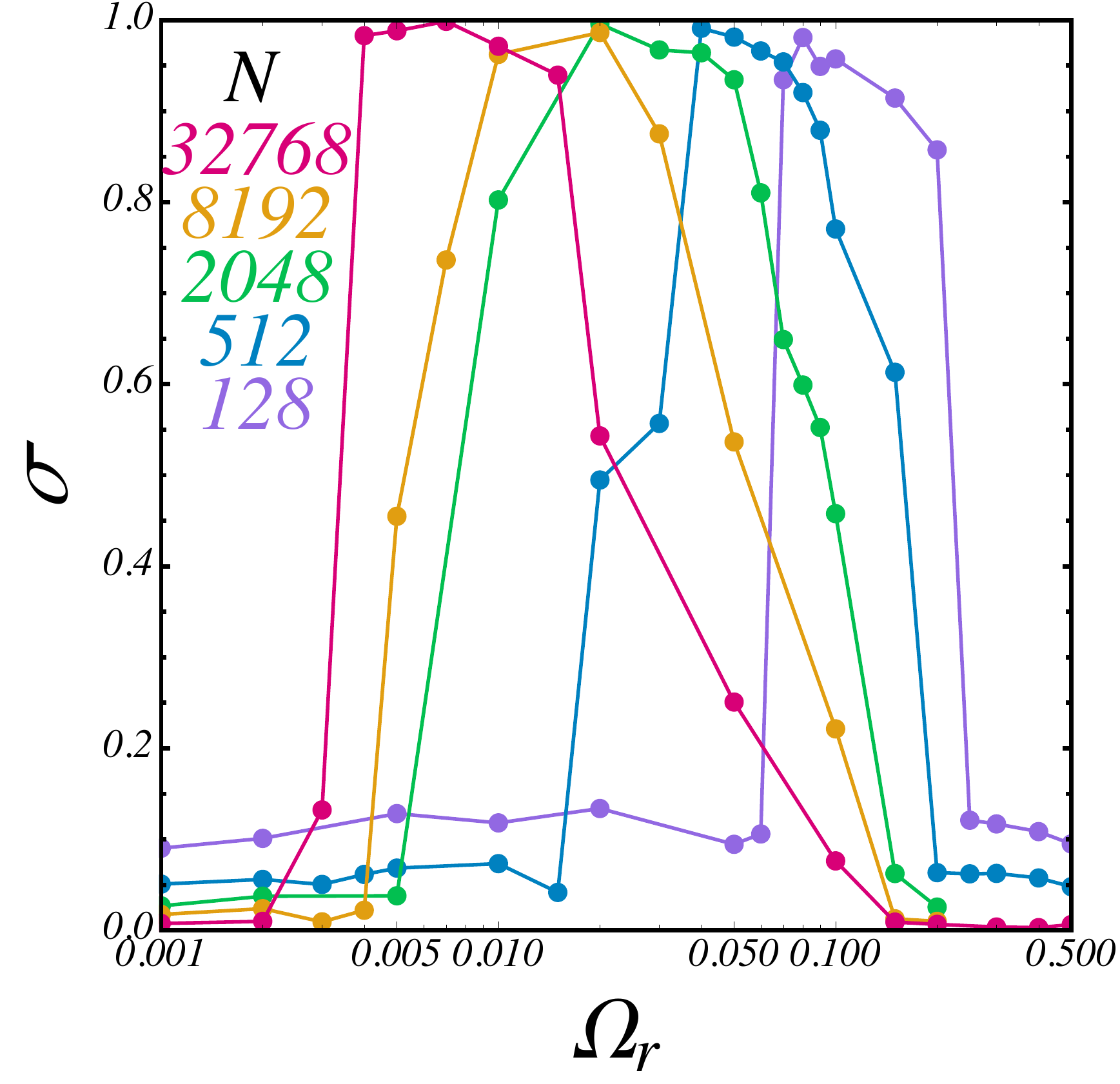}
    \caption{\textbf{Raw synchronization against relaxation.}
    Synchronization amplitude $\sigma$ against the raw relaxation rate $\Omega_r$ at $\phi =0.2$ at several system sizes, averaged over 10 realisations.
    }
    \label{fig:nonrescaledsigmas}
\end{figure}
In the main text, we show that the onset of the synchronization amplitude $\sigma$ in noiseless HABPs is located at a value of the relaxation rate $\Omega_r$ that scales like $L^{-1}$, resorting to plots of $\sigma$ against $\Omega_r L$.
In Fig.~\ref{fig:nonrescaledsigmas}, we show a raw set of curves of $\sigma$ against $\Omega_r$, that was used to plot Fig.~2$(a)$ of the main text.
As expected, the value of $\Omega_r$ that corresponds to the onset of $\sigma$ on the left-hand side is shifted to lower values as the size of the system increases, while the decay of $\sigma$ at high relaxation rates essentially always happens at the same relaxation rate, as it is set by the ratio between the size of orbits and that of targets, $R_0/a = 2/ (\pi \Omega_r)$, which does not depend on $L$.

\section{Synchronization with bidisperse relaxation rates\label{app:BinaryMixture}}

\begin{figure}
    \centering
    \includegraphics[width=0.2\textwidth]{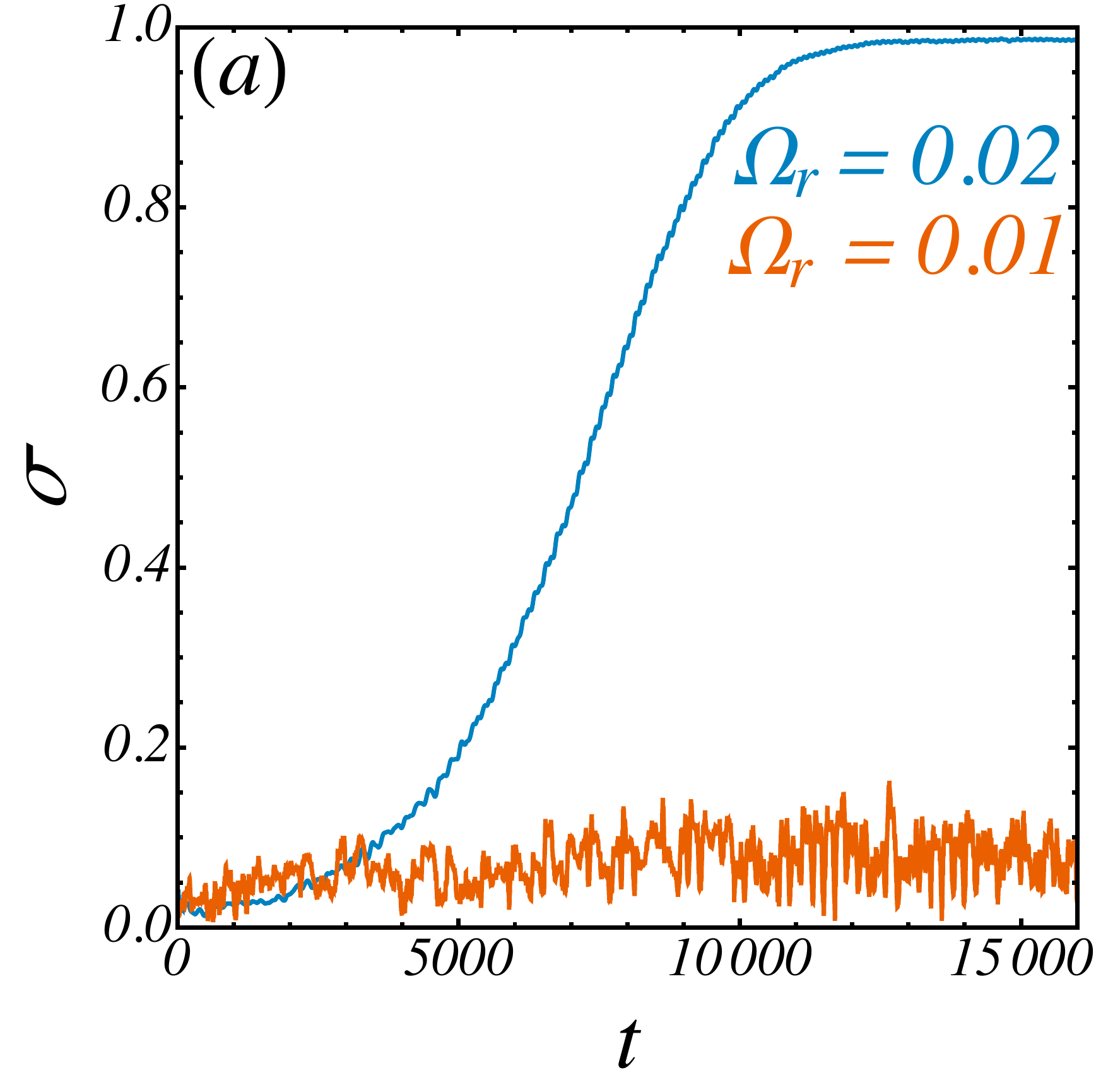}
    \includegraphics[width=0.2\textwidth]{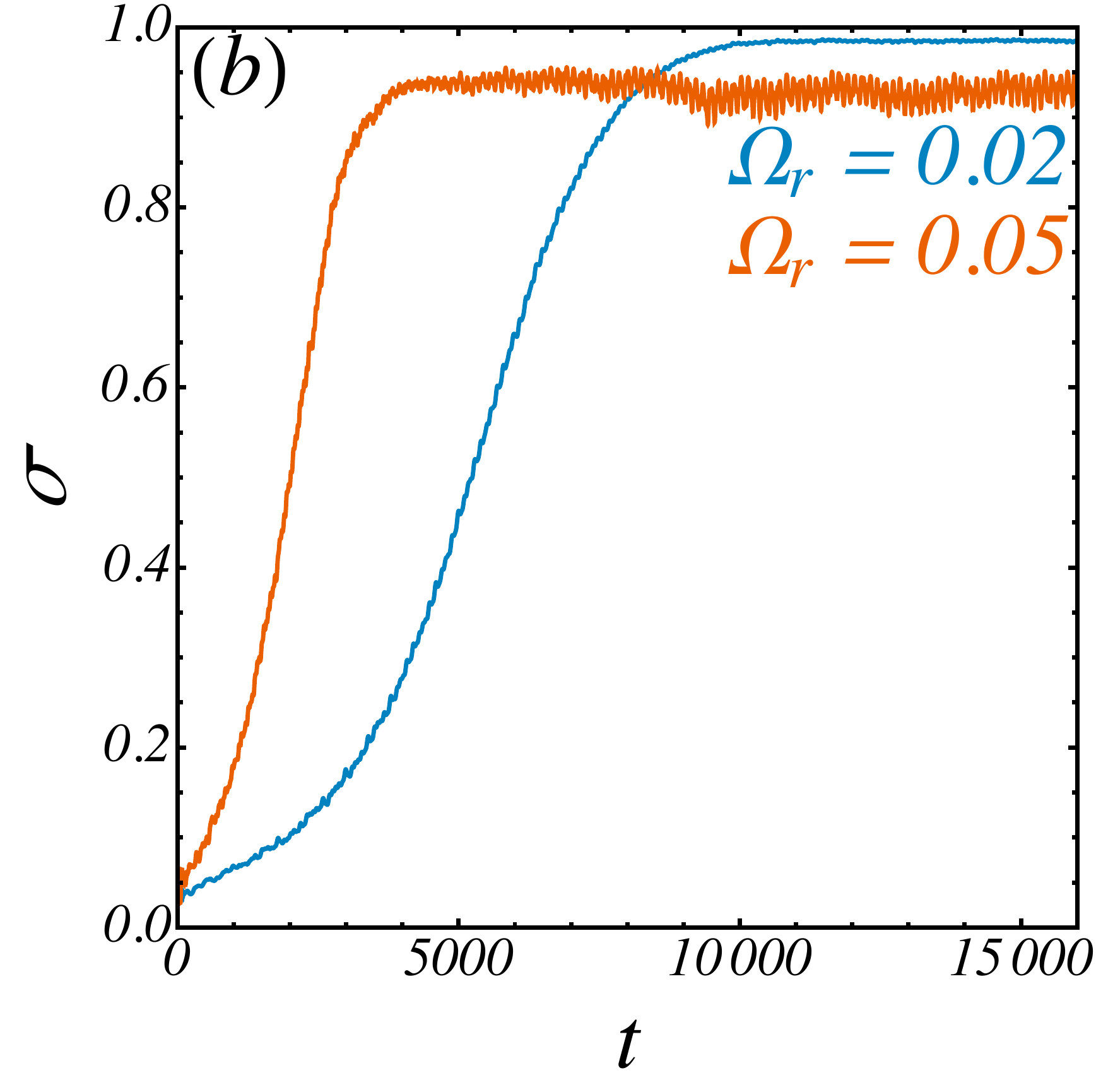}
    \includegraphics[width=0.2\textwidth]{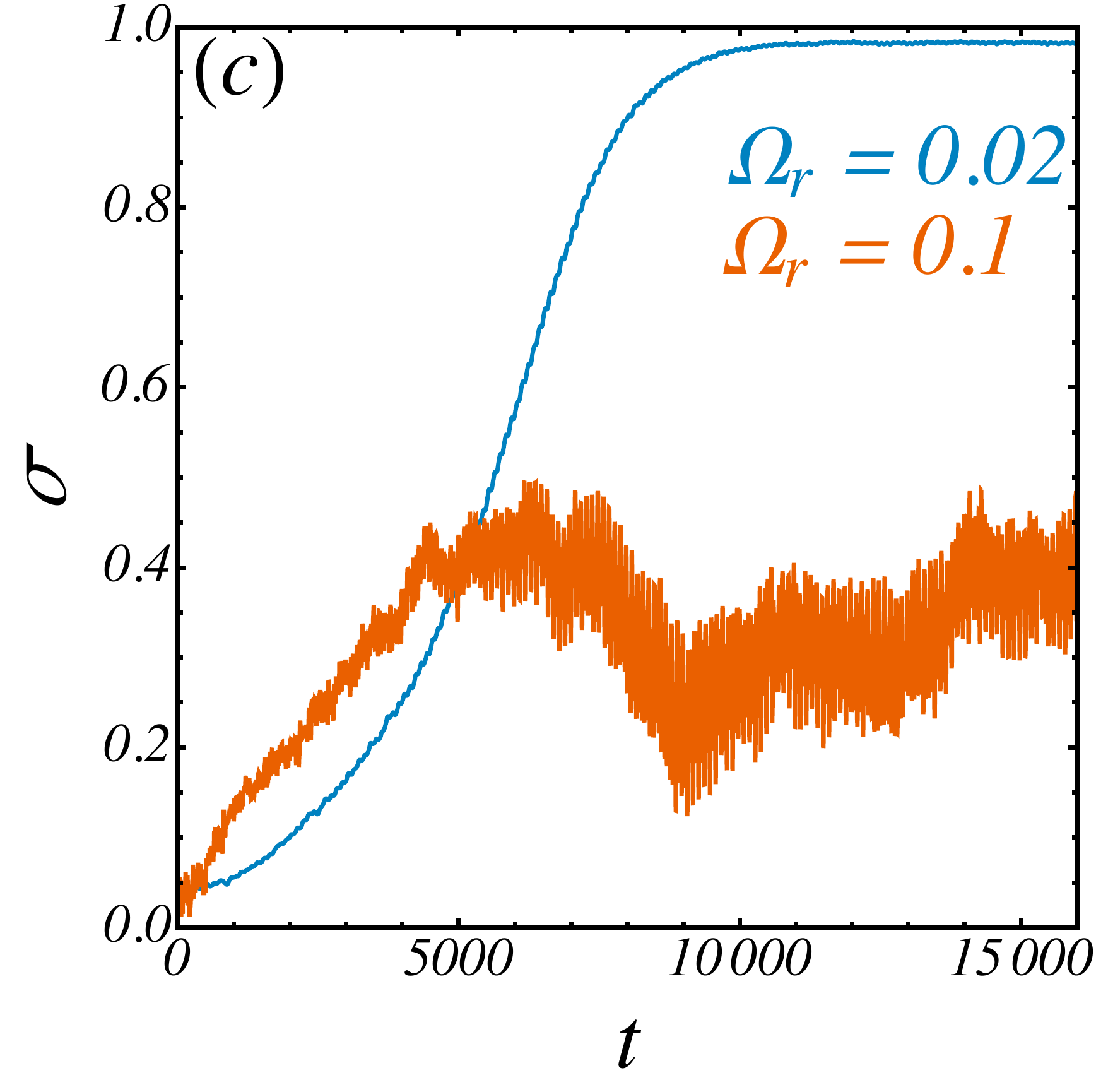}
    \includegraphics[width=0.2\textwidth]{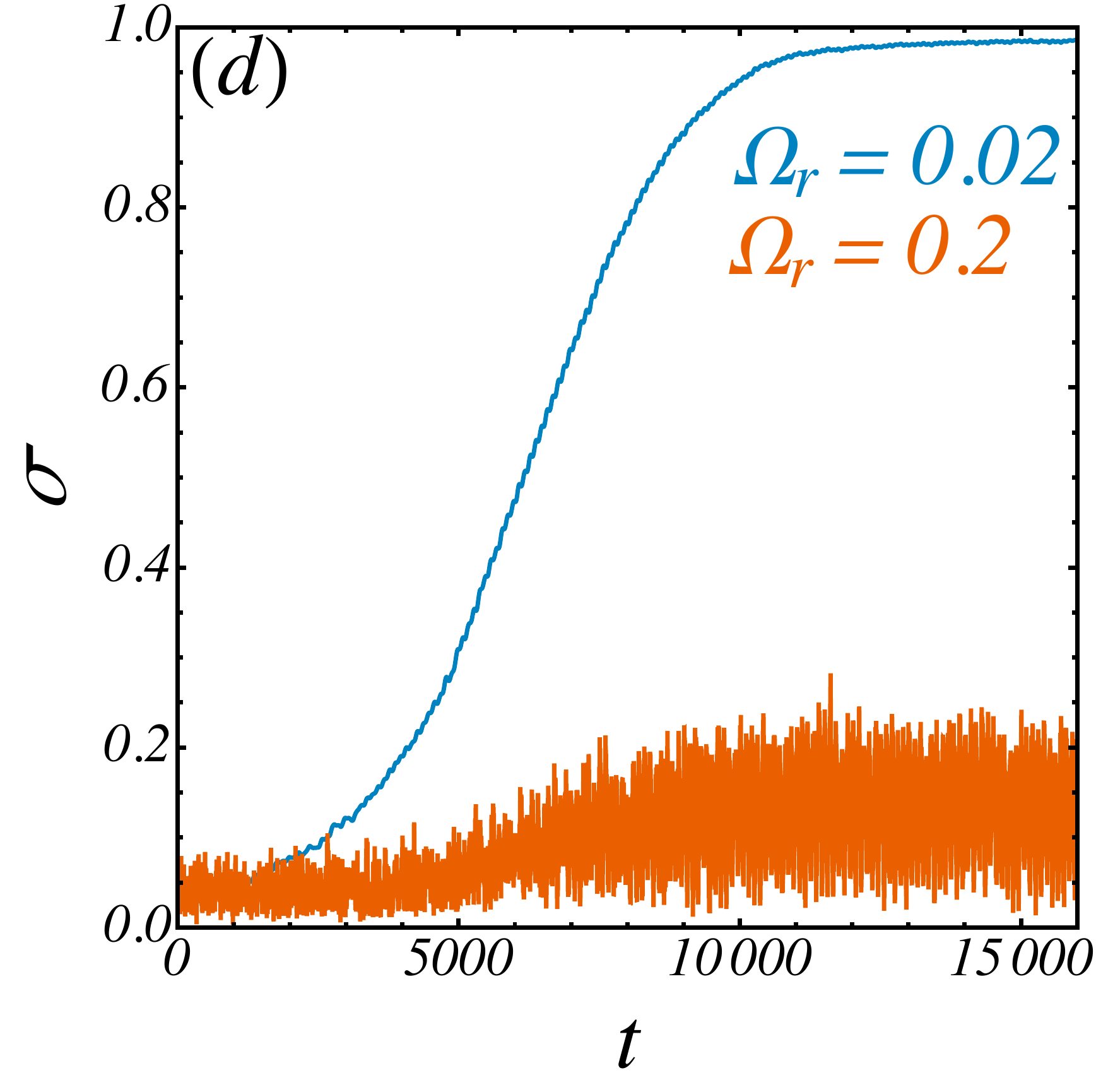}
    \caption{\textbf{Synchronization of bidisperse HABPs.}
    In each panel, we present the synchronization $\sigma_1$ of the group of particles with $\Omega_{r,1} = 0.02$ (blue) and $\sigma_2$ that of the group of particles with a relaxation rate $\Omega_{r,2}$ (orange), that is varied across panels.
    The values of $\Omega_{r,2}$ across $(a)-(d)$ are respectively $0.01$ (no orbits), $0.05, 0.1$ (smaller orbits) and $0.2$ (no orbits).
    }
    \label{fig:Bidisperse}
\end{figure}

In the main text, we mention the resistance of synchronization to mixing different values of the relaxation rate $\Omega_r$.
Here, we present additional data on the synchronization amplitude of systems set up as follows.
We consider a system of $N=2048$ HABPs, half of which evolve with a relaxation rate $\Omega_{r,1} = 0.02$, which yields a synchronization amplitude $\sigma \approx 1$ in the monodisperse case, while the other half evolves with $\Omega_{r,2}$ that is varied.
The two families of particles can therefore admit stable orbit solutions with different sizes and, since the self-propulsion speed is fixed, different periods: as a result, they cannot synchronize with each other.
Starting from uniformly drawn initial positions and orientations, we let the system evolve following the same dynamics as in the rest of this work.
During these dynamics, we record the synchronizations $\sigma_{1,2}$ of each family.
We present the results in Fig.~\ref{fig:Bidisperse}.
We show that, regardless of the value of $\Omega_{r,2}$, $\sigma_1 \approx 1$ as in the monodisperse case.
In other words, in spite of the collisions with a system that has different orbit sizes or, sometimes, doesn't even have orbits, the subsystem at $\Omega_{r,1}$ manages to self-organize into regular orbits and synchronize.
Furthermore, the synchronization of the second group, $\sigma_2$, also reaches the value that it is expected to have in the monodisperse case.
Videos in the SI show dynamics in both the $(\sigma_1 \approx 1, \sigma_2 \approx 0)$ and the $(\sigma_1 \approx 1, \sigma_2 \approx 1)$ cases.
These results show that the synchronization of HABPs is not a trivial consequence of the existence of a single orbit size: it is in fact robust to some degree of polydispersity.

\section{Single-particle Dynamics\label{app:SingleParticle}}

\subsection{Deterministic Case}

In the main text, we discuss the nature of the steady state of $N$ HABPs at a finite density, and describe the occurrence of large synchronised orbits.
These orbits can be observed because they are in fact steady states of the deterministic single-particle dynamics.
We here briefly discuss these dynamics.
They are described by the equations of motion of one particle,
\begin{align}
    \dot{\bm{r}} &= v_0 \hat{\bm{e}}(\theta), \\
    \dot{\theta} &= \omega_r (\theta_T - \theta),
\end{align}
and by the condition that a target absorbs the particle when they are at a distance $a$.
Without any loss of generality, one can place the target at the origin, so that at all times $\theta_T = \text{atan}(-y / x)$, and rescale time and space units by, respectively, $a/v_0$ and $v_0$.
The only free parameter in the equations of motion is then the non-dimensional relaxation rate $\Omega_r = \omega_r a / v_0$, to which one can associate the radius of the stable circular orbit allowed by the dynamics, $R_0 = 2 a/ (\pi \Omega_r)$.
At any given value of this parameter, one can test various initial conditions and record the long-time state of the system.

Using the rotational symmetry of the system, the choice of initial conditions can be reduced to that of the initial distance to the origin, $d_0 = \sqrt{x(0)^2 + y(0)^2}$, and of the initial orientation of self-propulsion $\theta_0 = \theta(0)$.
Since we enforce absorption by the target at a distance $a$, only distances $d_0 > a$ should be considered.
Furthermore, by symmetry, $\theta_0$ and $\theta_0 + \pi$ are bound to yield the same final state.
We therefore only need to explore the observed states of the particle for $d_0 \in \left] a; \infty\right]$ and $\theta_0 \in \left[ 0; \pi\right]$.

\begin{figure}[h]
    \centering
    \includegraphics[width=0.50\textwidth]{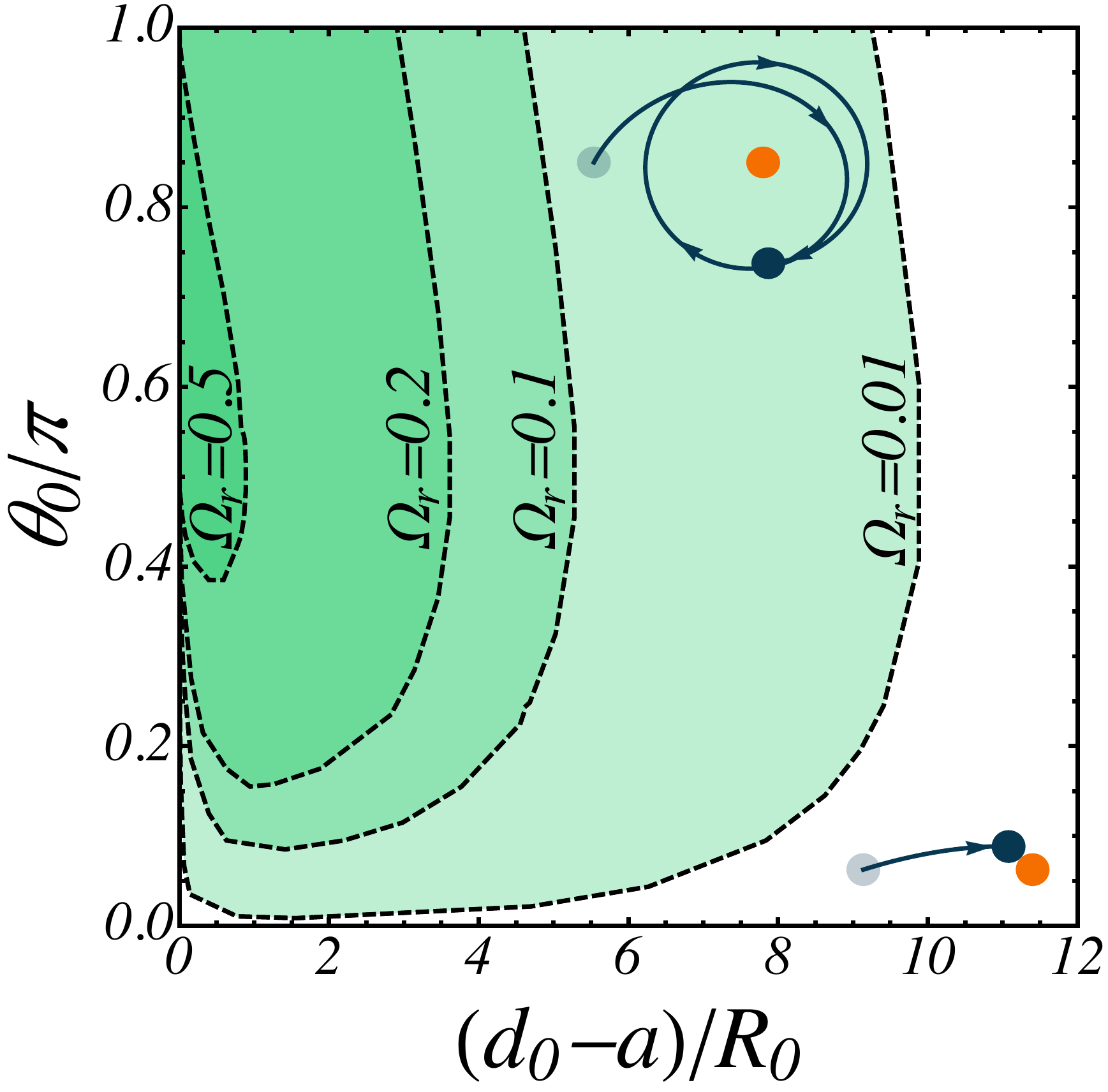}
    \caption{\textbf{Deterministic single-particle dynamics.}
    We plot (dashed black lines) the limiting lines between numerical observations of circular orbits (green domain) and absorption by the target (white domain) for a few values of $\Omega_r$, in the dimensionless initial distance to absorption $(d_0 -a)/R_0$ - initial self-propulsion orientation $\theta_0/\pi$ plane.
    The two kinds of observed trajectories are shown as insets.
    In these insets, we show the initial position of the particle as a transparent blue disk, its position when it enters the absorbing state in dark blue, the trajectory between the two as an arrow, and the position of the target as an orange disk.
    }
    \label{fig:SingleParticle}
\end{figure}
Using numerical integration of the dynamics, we find that only two kinds of trajectories are observed: some end up in an absorption by the target, while others end up on stable circular orbits at $R_0$.
These domains are mapped out for a few values of $\Omega_r$, in the $\left((d_0 - a)/R_0, \theta_0\right)$ plane Fig.~\ref{fig:SingleParticle}.
We show that, for values of $\Omega_r$ small enough ($\Omega_r < 2/pi \approx 0.6$) that orbits verify $R_0 > a$, there is a finite domain of initial conditions that leads to stable orbits at long times.
This domain is finite in both directions: there always exists a $d_0$ large enough that orbits are never observed regardless of $\theta_0$, and a $\theta_0$ small enough that no orbits are observed regardless of $d_0$.
Note that the value of $\theta_0$ for which the domain of orbits is the broadest is $\theta_0 = \pi/2$.
As $\Omega_r$ increases ($R_0 \to a$), the domain in which stable circular orbits are observed becomes smaller and smaller, until it essentially contains only the line $(d_0 = a, \pi/2 \leq \theta_0 \leq \pi)$.

The integration was here performed in free space: in the simulations of the main text there is the additional scale of the size of the periodic box that plays a role.
Essentially, if $2 R_0 < L < d_{max}$, meaning that the box can hold a full orbit but that it is smaller than maximal distance at which orbits are observed, one can expect stable orbits at all allowed distances, given the right $\theta_0$.
This does not, however, guarantee the stability of orbits in the many-body case.

\subsection{Noisy case}

One can also discuss the properties of a single HABP in the presence of noise terms.
In order to do so, let us first write the full Langevin equations of motion for a single particle tagged $i$, with both noise terms enabled,
\begin{align}
    \dot{\bm{r}}_i &= v_0 \hat{\bm{e}}_i(\theta) + \sqrt{2 D_T} \bm{\eta}_i\\
    \dot{\theta}_i &= \omega_r \left(\theta_{i,T} - \theta_i\right) + \sqrt{2 D_r} \xi_i,
\end{align}
where the difference $\left(\theta_{i,T} - \theta_i\right)$ should always be understood modulo $2\pi$.
Let us then define the probability density of finding that particle at position $\bm{r}$ and with orientation $\theta$ at time $t$,
\begin{align}
    p_i(\bm{r},\theta,t) = \delta\left( \bm{r}_i(t) - \bm{r} \right)\delta\left( \theta_i(t) - \theta \right),
\end{align}
where the probability should here be understood as a probability over realizations of the noise, and the probability $p$ to find any of $N$ independent particles around $\bm{r},\theta$ at time $t$, defined by
\begin{align}
    p(\bm{r},\theta,t) = \sum\limits_{i=1}^N p_i(\bm{r},\theta,t).
\end{align}
The last sum can be understood as a sum over independent runs with different initial conditions and different noise histories.
Consider a microscopic, one-particle observable $A$ that only depends on the coordinates of the particles. 
By definition, its value for particle $i$ can be written as
\begin{align}
    A(\bm{r}_i,\theta_i) &= \int d\bm{r} d\theta p_i(\bm{r},\theta,t) A(\bm{r},t). \label{eq:defObs}
\end{align}
One can apply the same equality to the observable $B = dA/dt$, leading to
\begin{align}
    \frac{dA}{dt}(\bm{r}_i,\theta_i) &= \int d\bm{r} d\theta p_i(\bm{r},\theta,t) \frac{dA}{dt}(\bm{r},t).
\end{align}
Since $A$ is only a function of two random variables that each follow overdamped Langevin equations, its time derivative can be written following the It\={o} convention, so that
\begin{align}
    \frac{dA}{dt}(\bm{r}_i,\theta_i) &= \int d\bm{r} d\theta p_i(\bm{r},\theta,t) \left[ \dot{\bm{r}}_i \nabla_{\bm{r}} A + \dot{\theta}_i \partial_\theta A + D_T \nabla_{\bm{r}}^2 A + D_r \partial_\theta^2 A \right].
\end{align}
Using the equations of motion, this equation becomes
\begin{align}
    \frac{dA}{dt}(\bm{r}_i,\theta_i) &= \int d\bm{r} d\theta p_i(\bm{r},\theta,t) \left[\left( v_0 \hat{\bm{e}}(\theta) + \sqrt{2 D_T} \bm{\eta}_i\right)\cdot\nabla_{\bm{r}} A + \left( \omega_r\left(\theta_{T,i} - \theta\right) + \sqrt{2 D_r} \xi_i \right) \partial_\theta A + D_T \nabla_{\bm{r}}^2 A + D_r \partial_\theta^2 A \right].
\end{align}
The derivatives can then be passed on to the probability field using integrations by parts, leading to 
\begin{align}
    \frac{dA}{dt}(\bm{r}_i,\theta_i) &= \int d\bm{r} d\theta A(\bm{r},\theta) \left[ - \nabla_{\bm{r}} \left\{ \left( v_0 \hat{\bm{e}}(\theta) + \sqrt{2 D_T} \bm{\eta}_i \right) p_i(\bm{r},\theta,t) \right\} + D_T \nabla_{\bm{r}}^2 p_i(\bm{r},\theta,t)  \right. \nonumber\\
     &- \left. \partial_\theta \left\{ \left( \omega_r\left(\theta_{T,i} - \theta\right) + \sqrt{2D_r} \xi_i \right) p_i(\bm{r},\theta,t)\right\}  + D_r \partial_\theta^2 p_i(\bm{r},\theta,t) \right].
\end{align}
Finally, one can notice that taking the time derivative of Eq.~\ref{eq:defObs} yields another expression for $dA/dt$,
\begin{align}
    \frac{dA}{dt}(\bm{r}_i,\theta_i) &= \int d\bm{r} d\theta \frac{\partial p_i}{\partial t}(\bm{r},\theta,t) A(\bm{r},\theta).
\end{align}
Since the equality between the two integrals is verified for any test function $A$, it implies the equality
\begin{align}
    \frac{\partial p_i}{\partial t}(\bm{r},\theta,t) &= - \nabla_{\bm{r}} \left\{ \left( v_0 \hat{\bm{e}}(\theta) + \sqrt{2 D_T} \bm{\eta}_i \right) p_i(\bm{r},\theta,t) \right\} + D_T \nabla_{\bm{r}}^2 p_i(\bm{r},\theta,t) \nonumber\\
    &- \partial_\theta \left\{ \left( \omega_r \left(\theta_{T,i} - \theta\right) + \sqrt{2 D_r} \xi_i \right) p_i(\bm{r},\theta,t)\right\}  + D_r \partial_\theta^2 p_i(\bm{r},\theta,t).
\end{align}
We finally sum the equations corresponding to all particles, and use the integral definition of the delta distribution to rewrite the sum of interaction forces as an integral, leading to
\begin{align}
    \frac{\partial p}{\partial t}(\bm{r},\theta,t) &= - \nabla_{\bm{r}} \left\{ \left( v_0 \hat{\bm{e}}(\theta) \right) p(\bm{r},\theta,t) \right\}  + D_T \nabla_{\bm{r}}^2 p(\bm{r},\theta,t) \nonumber \\
    & - \sum\limits_{i=1}^{N} \partial_\theta \left\{ \left( \omega_r\left(\theta_{T,i} - \theta\right) + \sqrt{2D_r} \xi_i  + \sqrt{2 D_T} \bm{\eta}_i \right) p_i(\bm{r},\theta,t)\right\}   + D_r \partial_\theta^2 p(\bm{r},\theta,t).
\end{align}
One can show that the noise term can readily be replaced by an equivalent, macroscopic white noise term that does not depend explicitly on the microscopic degrees of freedom, with the same statistical properties~\cite{Dean1996}, so that
\begin{align}
    \frac{\partial p}{\partial t}(\bm{r},\theta,t) &= - \nabla_{\bm{r}} \left\{ v_0 \hat{\bm{e}}(\theta) p(\bm{r},\theta,t) \right\} + D_T \nabla_{\bm{r}}^2 p(\bm{r},\theta,t) + \sqrt{2D_T} \nabla_{\bm{r}} \cdot \left[ \bm{\eta}\left(\bm{r},\theta,t\right)\sqrt{p(\bm{r},\theta,t)} \right] \nonumber\\
    &  - \sum\limits_{i=1}^{N} \partial_\theta \left\{ \left( \omega_r\left(\theta_{T,i} - \theta\right)_{per}\right) p_i(\bm{r},\theta,t)\right\} + D_r \partial_\theta^2 p(\bm{r},\theta,t)  + \sqrt{2D_r} \partial_\theta\left[ \xi\left(\bm{r},\theta,t\right)\sqrt{p(\bm{r},\theta,t)} \right], \label{eq:AlmostFP}
\end{align}
where, introducing the averaging $\langle \cdot \rangle$ over realizations of the noise,  
\begin{align}
    \left\langle \eta_a(\bm{r},\theta,t) \eta_b(\bm{r}',\theta',t')  \right\rangle &= \delta_{ab}\delta(\bm{r}-\bm{r}')\delta(\theta - \theta')\delta(t - t'). \\
    \left\langle \xi(\bm{r},\theta,t) \xi(\bm{r}',\theta',t')  \right\rangle &= \delta(\bm{r}-\bm{r}')\delta(\theta - \theta')\delta(t - t').
\end{align}
If one considers interacting particles, the remaining sum in Eq.~\ref{eq:AlmostFP} \textit{cannot} easily be coarse-grained into a single macroscopic field, as each particle is coupled to a \textit{single} target, not the field of all targets.
In the case of independent particles however, targets can all be placed at the same position without any loss of generality, and one finally gets the closed-form Fokker-Planck equation
\begin{align}
    \partial_t p(\bm{r},\theta, t) &= - v_0 \hat{\bm{e}}(\theta) \bm{\nabla}_{\bm{r}}\cdot\left[p(\bm{r},\theta, t)  \right]  + D_T \nabla_{\bm{r}}^2 p(\bm{r},\theta,t) + \sqrt{2D_T} \nabla_{\bm{r}} \cdot \left[ \bm{\eta}\sqrt{p(\bm{r},\theta,t)} \right]  \\
    & + \omega_r\partial_\theta\left[p(\bm{r},\theta, t)\left(\theta - \theta_T\right) \right] + D_r \partial_\theta^2 p(\bm{r},\theta, t) + \sqrt{2 D_r} \partial_\theta\left[ \xi \sqrt{p(\bm{r},\theta, t)} \right]. \label{eq:FPABPtarget}
\end{align}

It is then interesting to define some coarse-grained, mean density and polarization fields,
\begin{align}
    \rho(\bm{r},t) &= \left\langle \int d\theta p(\bm{r},\theta, t) \right\rangle,\\
    \bm{P}(\bm{r},t) &= \left\langle \int d\theta p(\bm{r},\theta, t) \hat{\bm{e}}(\theta) \right\rangle,
\end{align}
respectively.
Integrating Eq.~\ref{eq:FPABPtarget} over angles on the one hand, and multiplying it by $\hat{\bm{e}}(\theta)$, then integrating on the other hand, one gets, after averaging over realizations of the noise:
\begin{align}
    \partial_t \rho &= D_T \nabla_{\bm{r}}^2 \rho - v_0 \bm{\nabla}_{\bm{r}}\cdot\bm{P} \\
    \partial_t \bm{P} &= D_T \bm{\nabla}_{\bm{r}}^2 \bm{P} - \frac{v_0}{2} \bm{\nabla}_{\bm{r}} \rho - D_r \bm{P} + \omega_r\bm{R},
\end{align}
with $\bm{R}$ the field defined by
\begin{align}
    \bm{R}(\bm{r},t) = \left\langle \int d\theta \hat{\bm{e}}(\theta) \partial_\theta\left[ p(\bm{r},\theta,t) \left(\theta - \theta_T\right) \right]\right\rangle,
\end{align}
which, after an integration by parts, yields
\begin{align}
        \bm{R}(\bm{r},t) = \left\langle \int d\theta \hat{\bm{e}}(\theta - \frac{\pi}{2}) p(\bm{r},\theta,t) \left(\theta - \theta_T\right) \right\rangle,
\end{align}
This field encodes the effect on $p$ of the torque caused by the homing interaction to the target.
Assuming that the time scales for rotational diffusion and advection of the macroscopic fields are well separated, so that rotational dynamics are faster than the translational ones (small $Pe_r$), one can assume that the second equation is only relevant in steady state, so that
\begin{align}
   D_r \bm{P} &= D_T \bm{\nabla}_{\bm{r}}^2 \bm{P} - \frac{v_0}{2} \bm{\nabla}_{\bm{r}} \rho  + \omega_r \bm{R}.
\end{align}
Substituting this equation in the first one yields
\begin{align}
    \partial_t \rho &= (D_T + \frac{v_0^2}{2 D_r}) \bm{\nabla}_{\bm{r}}^2\rho - \frac{v_0 D_T}{D_r}\bm{\nabla}_{\bm{r}}^2 \bm{P} - \frac{v_0 \omega_r}{D_r} \bm{\nabla}_{\bm{r}}\cdot\bm{R}.
\end{align}
We henceforth assume that the density and polarisation fields are smooth enough in space that the Laplacian of $\bm{P}$ can be neglected.
In the equation above, one can substitute $\bm{P}$ using its steady-state expression to realise that this is tantamount to neglecting terms of order equal to or higher than $\nabla^3$ in the gradient expansion of $\rho$.
Doing so leads to 
\begin{align}
    \partial_t \rho &= (D_T + \frac{v_0^2}{2 D_r}) \bm{\nabla}_{\bm{r}}^2\rho - \frac{v_0 \omega_r}{D_r} \bm{\nabla}_{\bm{r}}\cdot\bm{R}.
\end{align}
Notice that this equation, as expected from the microscopic dynamics, simply yields diffusion with the effective diffusion coefficient observed for ABPs~\cite{Fily2012}, $D_{eff} = D_T + v_0^2/(2D_r)$, in the limit of $\omega_r \to 0$.

In order to understand the effects of being homing compared to simple ABP dynamics, we now focus on the $\bm{R}$ field.
To make it more tractable, we assume that the probability distribution function $p(\bm{r},\theta,t)$ is factorizable into $p(\bm{r},\theta,t) = \rho(\bm{r},t) \Upsilon(\theta)$, with $\Upsilon$ a steady-state distribution of polarities.
Since the homing interaction plays the same role as a magnetic field $h$ acting on an internal polarity, in analogy with equilibrium continuous spins~\cite{Kardar2007a}, we assume that $\Upsilon$ is a Von Mises distribution, with a parameter $\lambda$ to be determined that plays the role of $\beta h$ in a spin system at the inverse temperature $\beta$
\begin{align}
    \Upsilon(\theta) \equiv \frac{e^{\lambda \cos(\theta - \theta_T)}}{2 \pi I_0(\lambda)}.
\end{align}
Under this hypothesis, the integral at hand is 
\begin{align}
    \bm{R} = \frac{\rho}{2\pi I_0(\lambda)}\int\limits_{-\pi}^{\pi} d\theta e^{\lambda \cos(\theta - \theta_T)}(\theta - \theta_T)\hat{\bm{e}}(\theta - \frac{\pi}{2}).
\end{align}
It is convenient to switch to the variable $\vartheta = \theta - \theta_T$, which takes care of the periodicisation of the relaxation term, leading to
\begin{align}
    \bm{R} = \frac{\rho}{2\pi I_0(\lambda)}\int\limits_{-\pi}^{\pi} d\vartheta e^{\lambda \cos\vartheta}\vartheta\hat{\bm{e}}(\vartheta + \theta_T - \frac{\pi}{2}).
\end{align}
One can then rewrite
\begin{align}
    \hat{\bm{e}}(\vartheta + \theta_T - \frac{\pi}{2}) &= \sin\left(\vartheta + \theta_T\right)\hat{\bm{e}}_x - \cos\left(\vartheta + \theta_T\right)\hat{\bm{e}}_y
    \\ &= \left[\cos\vartheta \sin\theta_T + \sin\vartheta\cos\theta_T \right] \hat{\bm{e}}_x - \left[\cos\vartheta \cos\theta_T - \sin\vartheta\sin\theta_T \right] \hat{\bm{e}}_y,
\end{align}
and notice that, by parity,
\begin{align}
    \int\limits_{-\pi}^{\pi} d\vartheta e^{\omega_r \cos\vartheta /D_r}\vartheta \cos\vartheta = 0.
\end{align}
There is only one other integral left to compute,
\begin{align}
    \mathcal{I} &\equiv \int\limits_{-\pi}^{\pi} d\vartheta e^{\lambda \cos\vartheta}\vartheta \sin\vartheta.
\end{align}
An integration by parts yields
\begin{align}
    \mathcal{I} &= \left[ - \frac{\vartheta}{\lambda} e^{\lambda \cos\vartheta } \right]_{-\pi}^{\pi} + \int\limits_{-\pi}^{\pi} d\vartheta \frac{e^{\lambda \cos\vartheta}}{\lambda} \\
                &= \frac{2\pi}{\lambda} \left(I_0(\lambda) - e^{-\lambda} \right).
\end{align}
All in all, one gets an expression for $\bm{R}$,
\begin{align}
    \bm{R} = \frac{\rho }{\lambda} \left(1 - \frac{e^{-\lambda}}{I_0(\lambda)} \right) \bm{\hat{e}}\left(\theta_T\right).
\end{align}
For convenience, we define the short-hand notation
\begin{align}
    f(\lambda) \equiv \frac{1}{\lambda}\left(1 - \frac{e^{-\lambda}}{I_0(\lambda)} \right),
\end{align}
as this function depends on the precise choice of the distribution of angles, whereas the density factor and the vector that carries $\bm{R}$ are generic for any distribution that is symmetric around $\theta_T$.

The final expression of $\bm{R}$ can be injected into the equation on $\rho$, leading to
\begin{align}
    \partial_t \rho &= D_{eff} \bm{\nabla}_{\bm{r}}^2\rho - \frac{v_0 \omega_r}{D_r}  f(\lambda) \bm{\nabla}_{\bm{r}}\cdot\left(\rho  \hat{\bm{e}}(\theta_T)\right).
\end{align}
The nabla can then be applied to the product in the usual way, leading to 
\begin{align}
    \partial_t \rho &= D_{eff} \bm{\nabla}_{\bm{r}}^2\rho - \frac{v_0 \omega_r}{D_r} f(\lambda) \left[\hat{\bm{e}}(\theta_T) \cdot  \bm{\nabla}_{\bm{r}} \rho + \rho \bm{\nabla}_{\bm{r}}\cdot\hat{\bm{e}}(\theta_T)\right].
\end{align}
Using a chain rule for the last term, this equation can be rewritten as
\begin{align}
    \partial_t \rho &= D_{eff} \bm{\nabla}_{\bm{r}}^2\rho -  \frac{v_0 \omega_r}{D_r} f(\lambda) \left[\hat{\bm{e}}(\theta_T) \cdot  \bm{\nabla}_{\bm{r}} \rho + \rho \hat{\bm{e}}\left(\theta_T + \frac{\pi}{2}\right) \cdot \bm{\nabla}_{\bm{r}}\theta_T\right].
\end{align}
One can finally use the expression of the $\theta_T$ field,
\begin{align}
    \theta_T &= \arctan\left[\frac{y_T - y}{x_T -x}\right],
\end{align}
so that, in particular,
\begin{align}
    \bm{\nabla}_{\bm{r}} \theta_T &= \frac{y_T - y}{\left\|\bm{r}_T - \bm{r} \right\|^2} \hat{\bm{e}}_x- \frac{x_T - x}{\left\|\bm{r}_T - \bm{r} \right\|^2} \hat{\bm{e}}_y, \\
    \cos \theta_T &= \frac{x_T - x}{\left\|\bm{r}_T - \bm{r} \right\|},\\
    \sin \theta_T &= \frac{y_T - y}{\left\|\bm{r}_T - \bm{r} \right\|}.
\end{align}
Using these equations, one notices that 
\begin{align}
    \hat{\bm{e}}\left(\theta_T + \frac{\pi}{2}\right)\cdot \bm{\nabla}_{\bm{r}} \theta_T &= - \frac{1}{\left\|\bm{r}_T - \bm{r} \right\|}, \\
    \hat{\bm{e}}(\theta_T) \cdot  \bm{\nabla}_{\bm{r}} \rho &= \frac{x_T - x}{\left\|\bm{r}_T - \bm{r} \right\|} \partial_x \rho + \frac{y_T - y}{\left\|\bm{r}_T - \bm{r} \right\|}\partial_y \rho.
\end{align}
Since we are here studying a single-particle problem, the target can be placed at the origin, leading to the simpler equation
\begin{equation}
\partial_t \rho = D_{eff} \bm{\nabla}_{\bm{r}}^2\rho + \frac{v_0 \omega_r}{D_r} f(\lambda)\,\left[ \rho + \left(\bm{\nabla}_{\bm{r}} \rho\right)\cdot\bm{r}\right].
\end{equation}

In order to understand the role of noise on single-particle trajectories, we simply seek steady-state, radially symmetric solutions of this equation.
In other words, we seek solutions of the ODE
\begin{align}
    \frac{d^2}{d r^2}\rho + \frac{1}{r}\frac{d}{d r}\rho + \kappa \left(\frac{d}{d r} \rho + \frac{\rho}{r} \right) &= 0,
\end{align}
with the inverse length $\kappa$ defined through
\begin{align}
    \kappa = \frac{v_0 \omega_r}{D_r D_{eff}} f\left(\lambda\right),
\end{align}
and the boundary conditions
\begin{align}
    \rho(a) &= 0, \\
    \rho(d) &= \rho_0.
\end{align}
The first boundary condition encodes the fact that the target is an absorbing boundary at a distance $a$ (the targets' and particles' diameters), while the second one encodes the presence of a source of particles at a constant density at some distance $d$, that one can for instance picture as an initial distance from which particles are initially drawn.
This ODE is cast into a form that can be used as a definition for $Ei$, the exponential integral~\cite{Abramowitz1972}, so that the corresponding solutions can be written as
\begin{align}
    \rho(r) &= \mathds{1}\left(r\geq a\right) \rho_0 e^{\kappa\left(a - r\right)} \frac{ Ei(\kappa d) - Ei(\kappa r)}{Ei(\kappa d) - Ei(\kappa a)}. \label{eq:SSrho}
\end{align}
The choice of $\rho_0$ is then imposed by the condition that this density should integrate to 1 over the whole annulus, so that
\begin{align}
    \frac{1}{\rho_0}    &= 2 \pi \int_{a}^{d} dr\,r e^{\kappa\left(a - r\right)} \frac{ Ei(\kappa d) - Ei(\kappa r)}{Ei(\kappa d) - Ei(\kappa a)} \\
                        &= \frac{2\pi}{\kappa^2}\left( e^{\kappa d}\frac{ \ln(d/a) + \kappa (d-a)}{Ei(\kappa d) - Ei(\kappa a)} - \left( 1 + \kappa d\right) \right).
\end{align}
Note that these solutions are ill-defined in the limit $\kappa \to 0$, but that the ODE can easily be solved in the case $\kappa = 0$, in which they yield a logarithmic profile on the annulus, as expected from simple diffusion.

To use the steady-state solution found above, it is useful to rewrite the inverse length $\kappa$ in terms of the dimensionless quantities introduced in the main text.
After some algebra, one finds
\begin{align}
    \kappa = \frac{2}{a} \frac{Pe_r \Omega_r}{Pe_r + 2 / Pe} f (\lambda).
\end{align}
This expression is useful to determine a reasonable dependence of $\lambda$ on the other parameters.
Indeed, we expect to recover purely diffusive dynamics ($\kappa\to 0$) in three different limits: $\Omega_r \to 0$ (no relaxation to the target), $\Omega_r$ finite but $Pe_r \Omega_r \to 0$ (relaxation is obscured by rotational noise), and $Pe \to 0$ (self-propulsion is obscured by translational noise).

The function $f$ itself is a strictly decreasing function with limiting values $f(0) = 1$ and $f(\lambda) \to 0$ as $\lambda\to\infty$. 
Therefore, when proposing an expression of $\lambda(Pe,Pe_r,\Omega_r)$, one should ensure three conditions.
First, that $\lambda \to 0$ as $Pe \to 0$, so that $\kappa \propto Pe f(0) \to 0$ as well.
Second, that $\lambda \to \infty$ as $Pe \to \infty$ and $Pe_r \to 0$, so that $\kappa \propto f(\infty) \to 0$.
Finally, that $\kappa \to 0$ for any values of $Pe$ and $Pe_r$ when $\Omega_r \to 0$.

Seeing the dependence on the two Péclet numbers imposed by these conditions, we propose that $\lambda$ should be inversely proportional to the effective diffusion constant $D_{eff}$, that plays the role of an effective temperature in terms of alignment towards the target.
It should be compared to the effective spatial diffusion associated to angular relaxation.
Within a time $1/\omega_r$, a particle moves a distance $v0/\omega_r$, so that one can construct the quantity $(v_0/\omega_r)^2 (v_0/a) = v_0^3 / (a \omega_r^2)$, homogeneous to a diffusion constant, that $D_{eff}$ can be compared to.
In short, we posit that
\begin{align}
    \lambda = \frac{v_0^3}{ a \omega_r^2 D_{eff}} = \frac{2}{\Omega_r^2 \left(Pe_r + 2/Pe \right)}.
\end{align}
One can check that this expression satisfies all three conditions above, as the expression of $\kappa$ then reduces to
\begin{align}
    \kappa = \frac{1}{a} Pe_r \Omega_r^3 \left(1 - \frac{\exp\left[-\frac{2}{\Omega_r^2 \left(Pe_r + 2/Pe \right)}\right]}{I_0\left[\frac{2}{\Omega_r^2 \left(Pe_r + 2/Pe \right)}\right]}\right). \label{eq:kappadependence}
\end{align}
\begin{figure}
    \centering
    \includegraphics[width=0.42\textwidth]{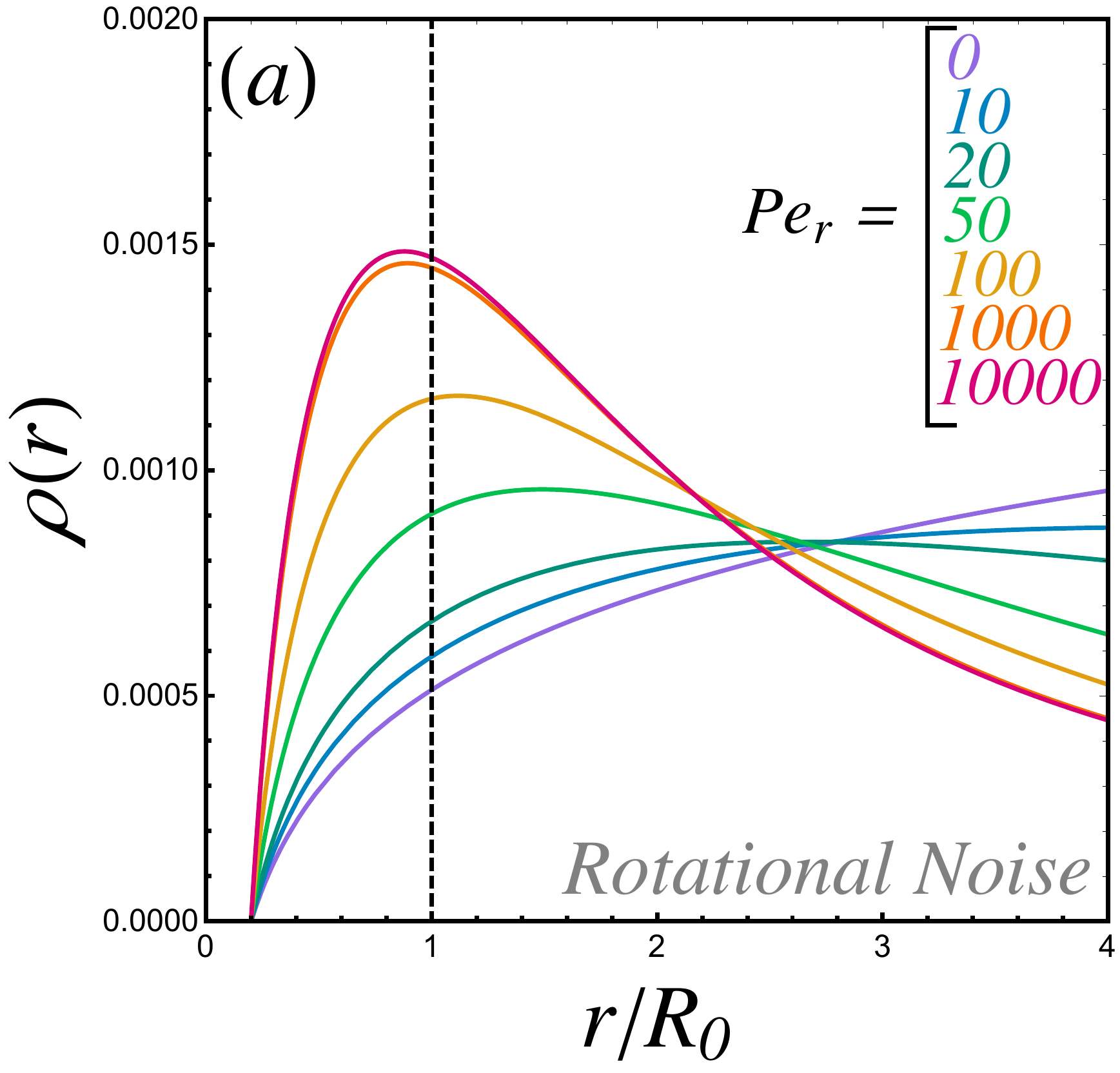}
    \includegraphics[width=0.42\textwidth]{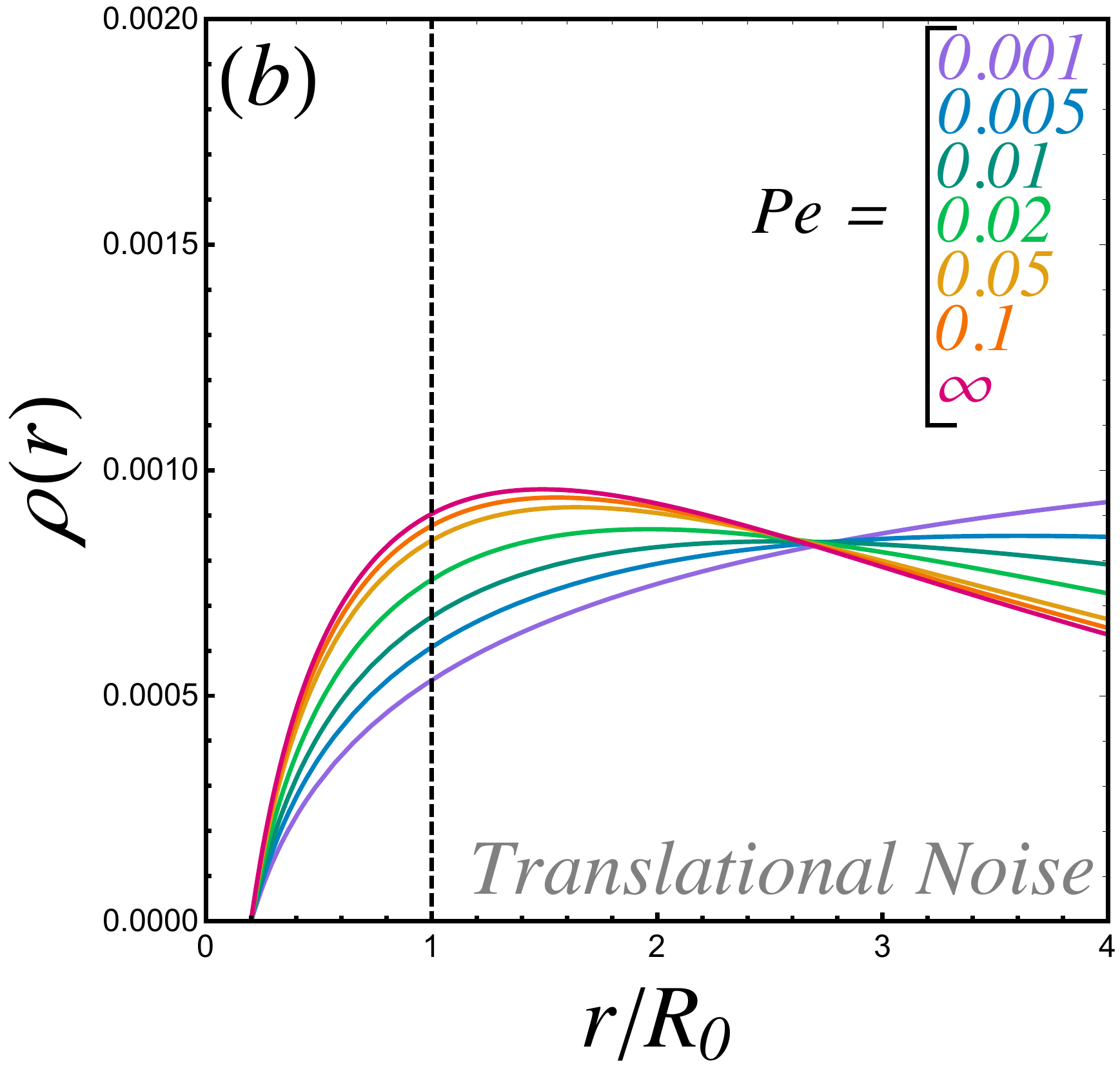}
    \caption{\textbf{Steady-state density profiles.}
    Curves obtained by plotting normalised profiles of Eq.~(\ref{eq:SSrho}) with $\kappa$ given by the ansatz of Eq.~(\ref{eq:kappadependence}).
    $(a)$ We set $Pe = \infty$ (no translational noise), $a=1$, $d=20$ and fix $\Omega_r = 2a/(\pi R_0)$ with an orbit radius $R_0 = 5$, then vary the rotational Péclet $Pe_r$. The noise amplitude grows from red to mauve.
    $(b)$ We set $Pe_r = 50$ and keep the other parameters unchanged.}
    \label{fig:NoisyParticle}
\end{figure}

We now check the shape of the density profiles given by Eq.~(\ref{eq:SSrho}), with the choice of $\kappa$ given by Eq.~(\ref{eq:kappadependence}).
The results are shown in Fig.~\ref{fig:NoisyParticle}.
In panel $(a)$, we set the translational noise to $0$ and vary the rotational Péclet number $Pe_r$ at a fixed value of the relaxation rate $\Omega_r$.
We show our simple approximation captures the appearance of a local maximum of the density at the expected orbit radius if the noise amplitude is low enough.
Beyond a critical value of the noise (that here depends on the arbitrary value of the outer distance $d$), this maximum disappears, and one eventually recovers the logarithmic density profile associated to pure diffusion.
In panel $(b)$, we show a similar set of curves obtained when varying the amplitude of the translational noise at a fixed value of the rotational noise (the full study above should be repeated in the case $D_r = 0$ strictly, since that value forbids the substitution of $\bm{P}$ into the equation on $\rho$ that we used at the start).

\begin{figure}
    \centering
    \includegraphics[width=0.18\textwidth]{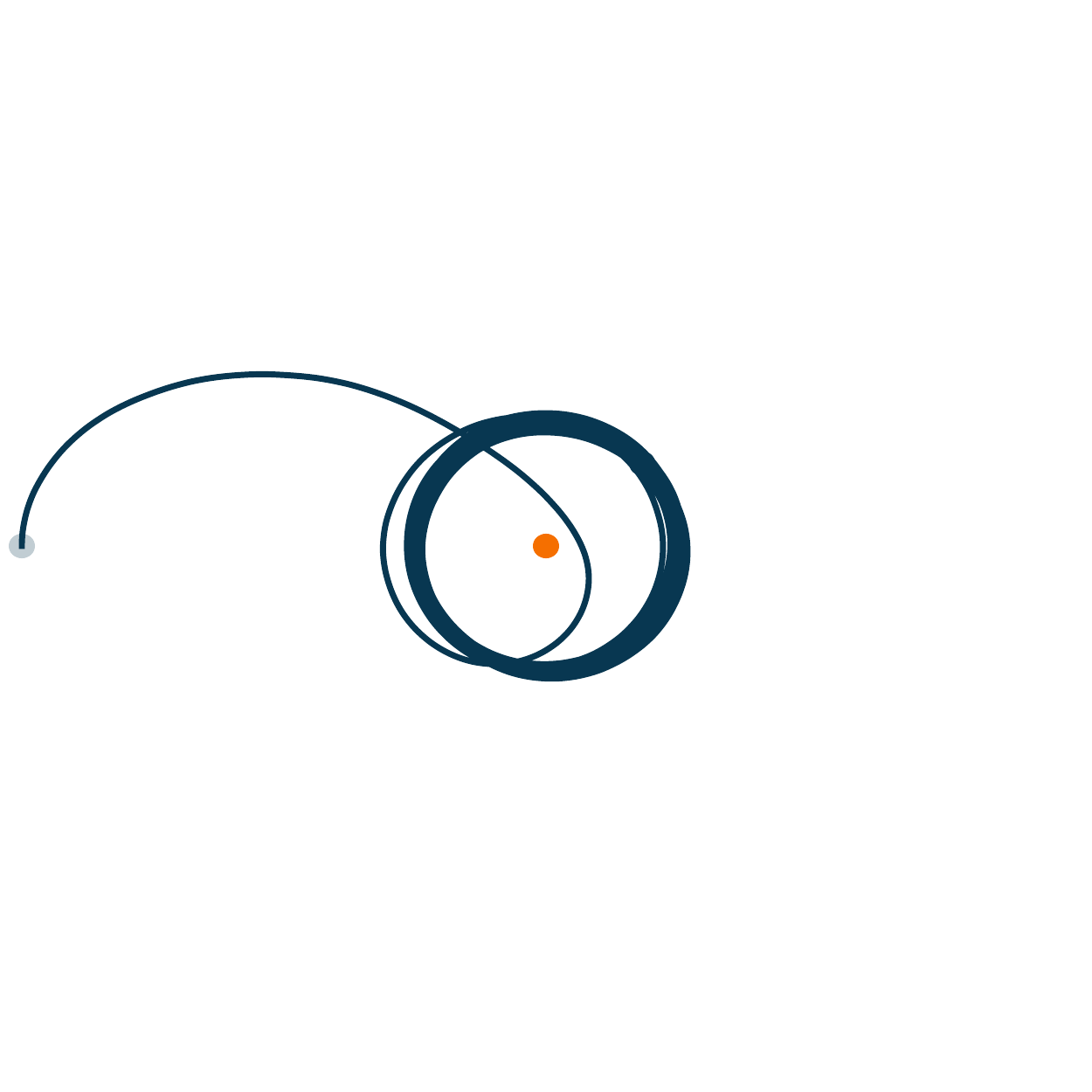}
    \includegraphics[width=0.18\textwidth]{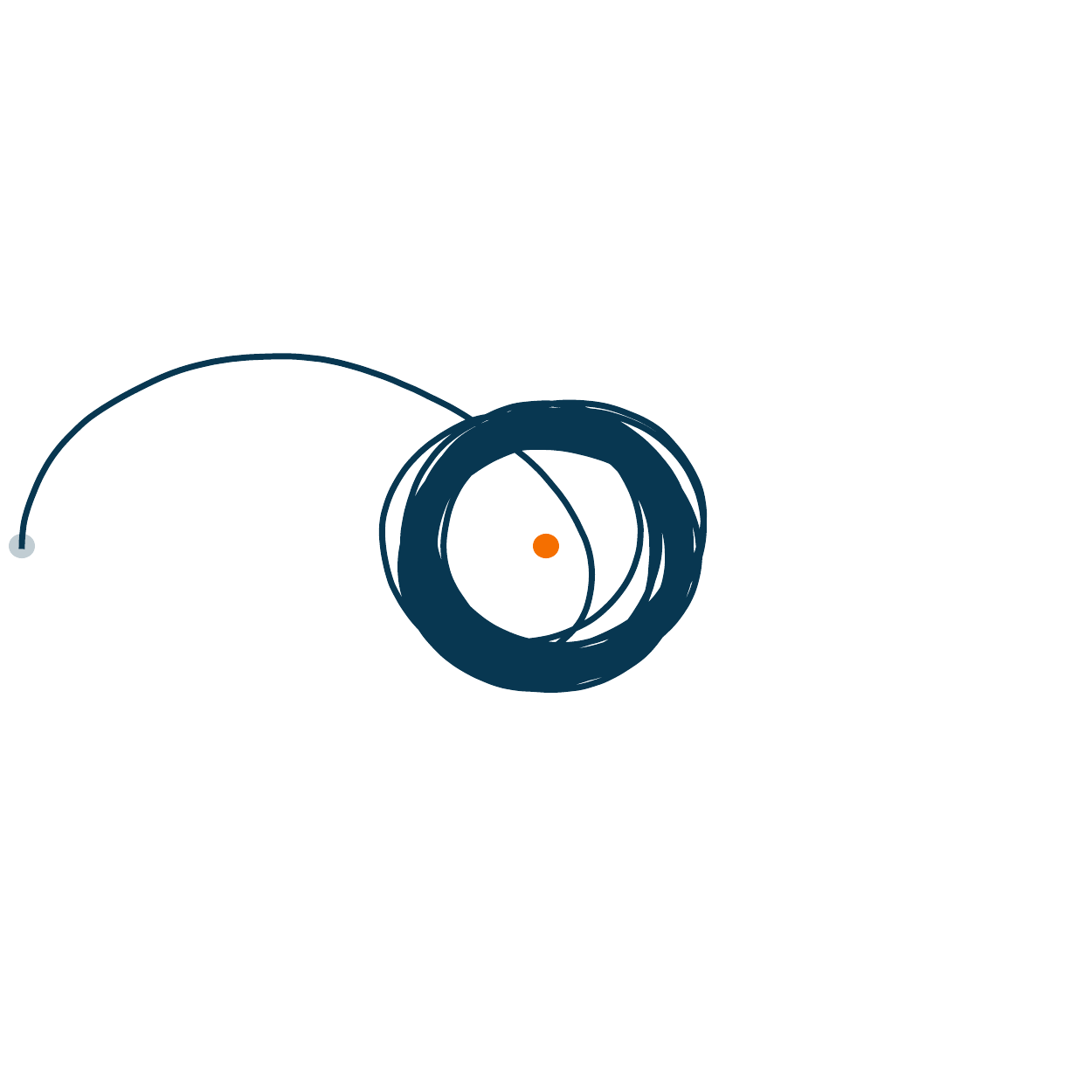}
    \includegraphics[width=0.18\textwidth]{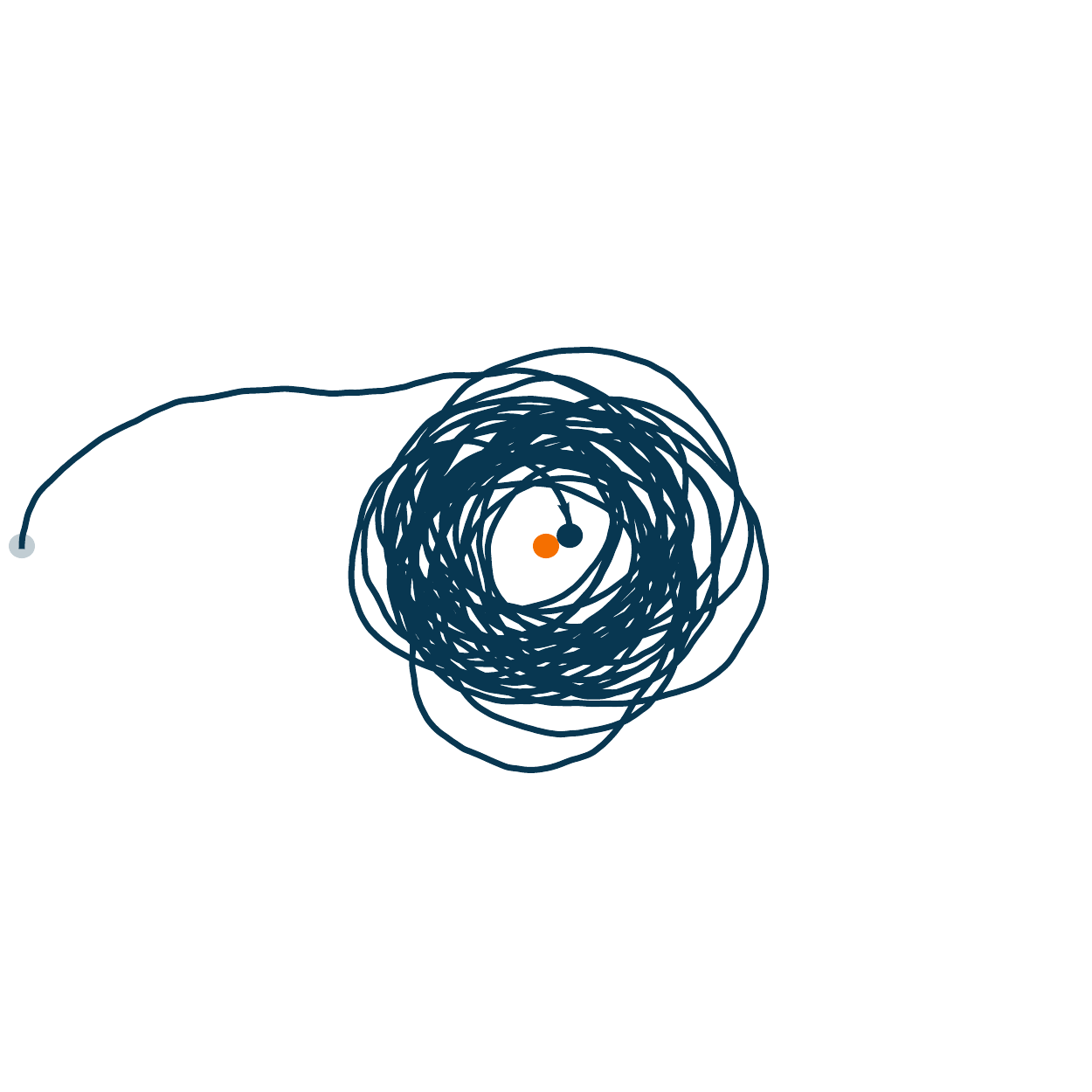}
    \includegraphics[width=0.18\textwidth]{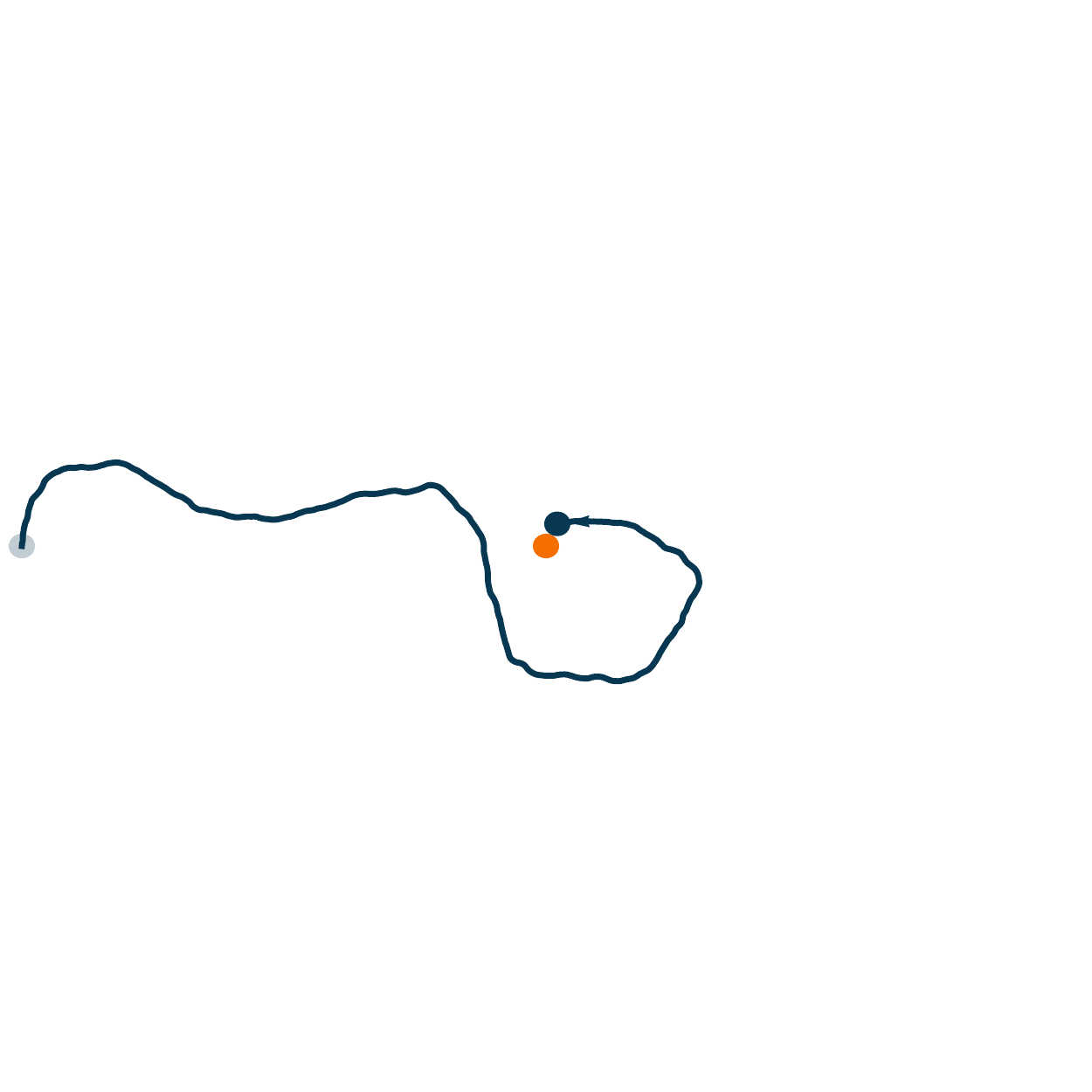}
    \includegraphics[width=0.18\textwidth]{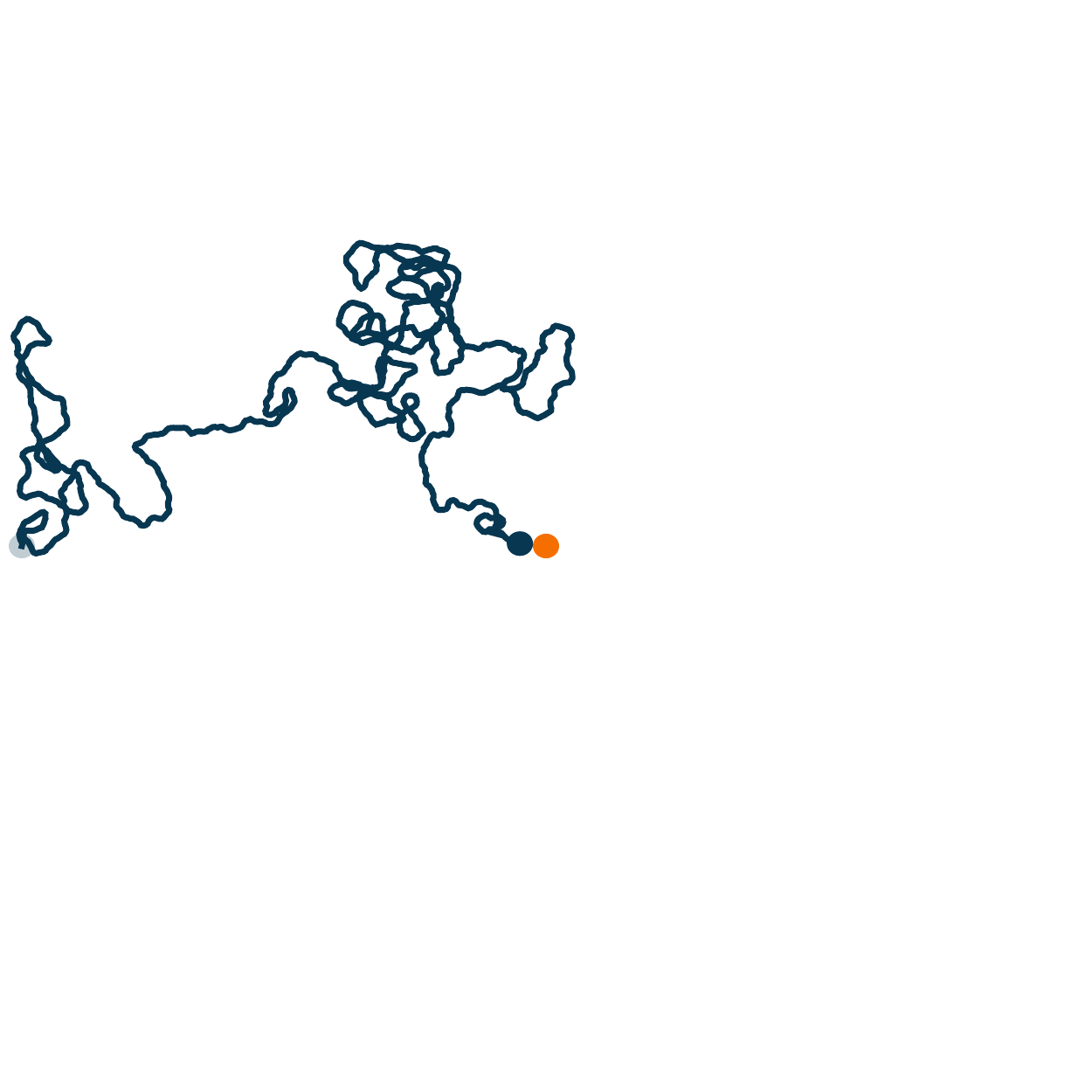} \\
    \includegraphics[width=0.18\textwidth]{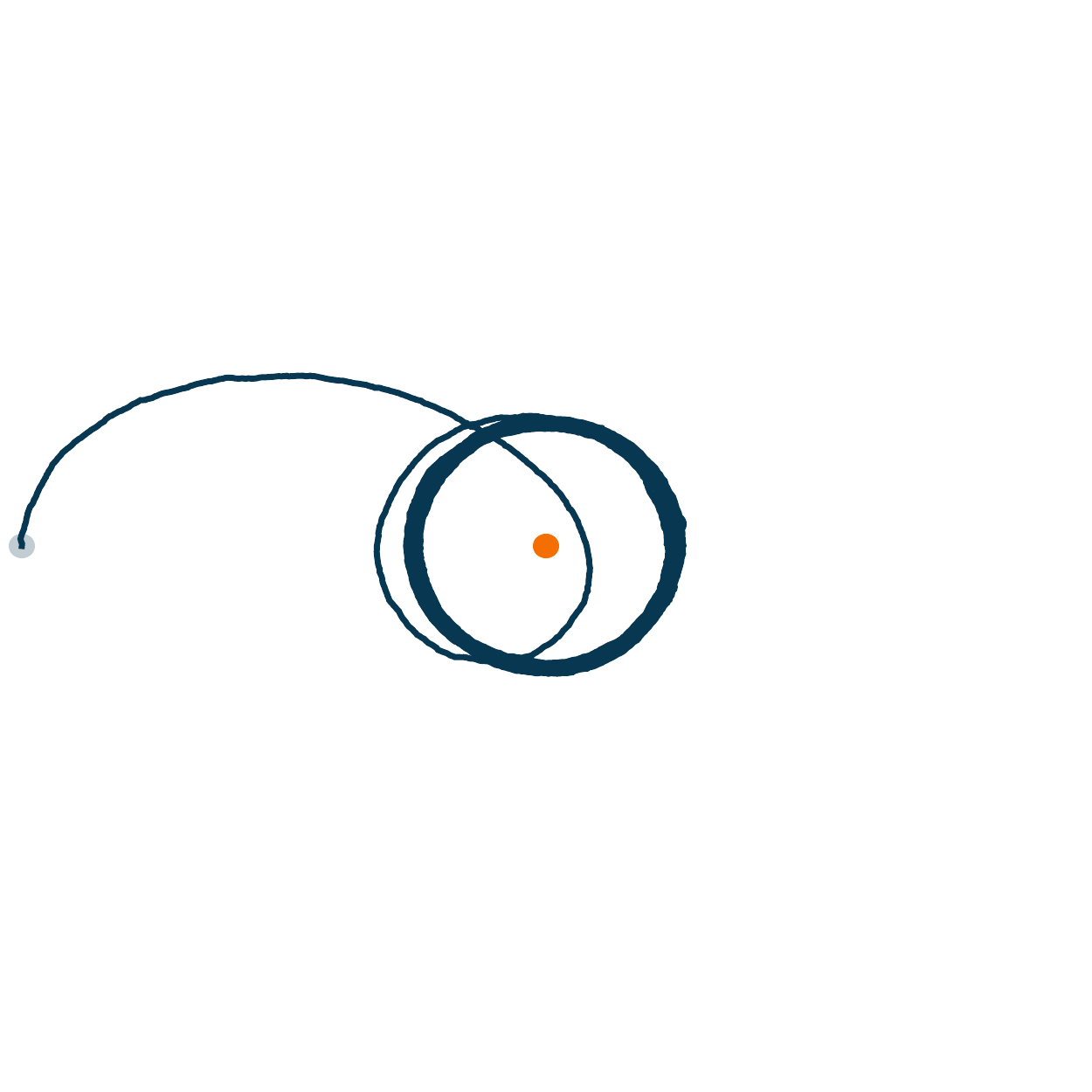}
    \includegraphics[width=0.18\textwidth]{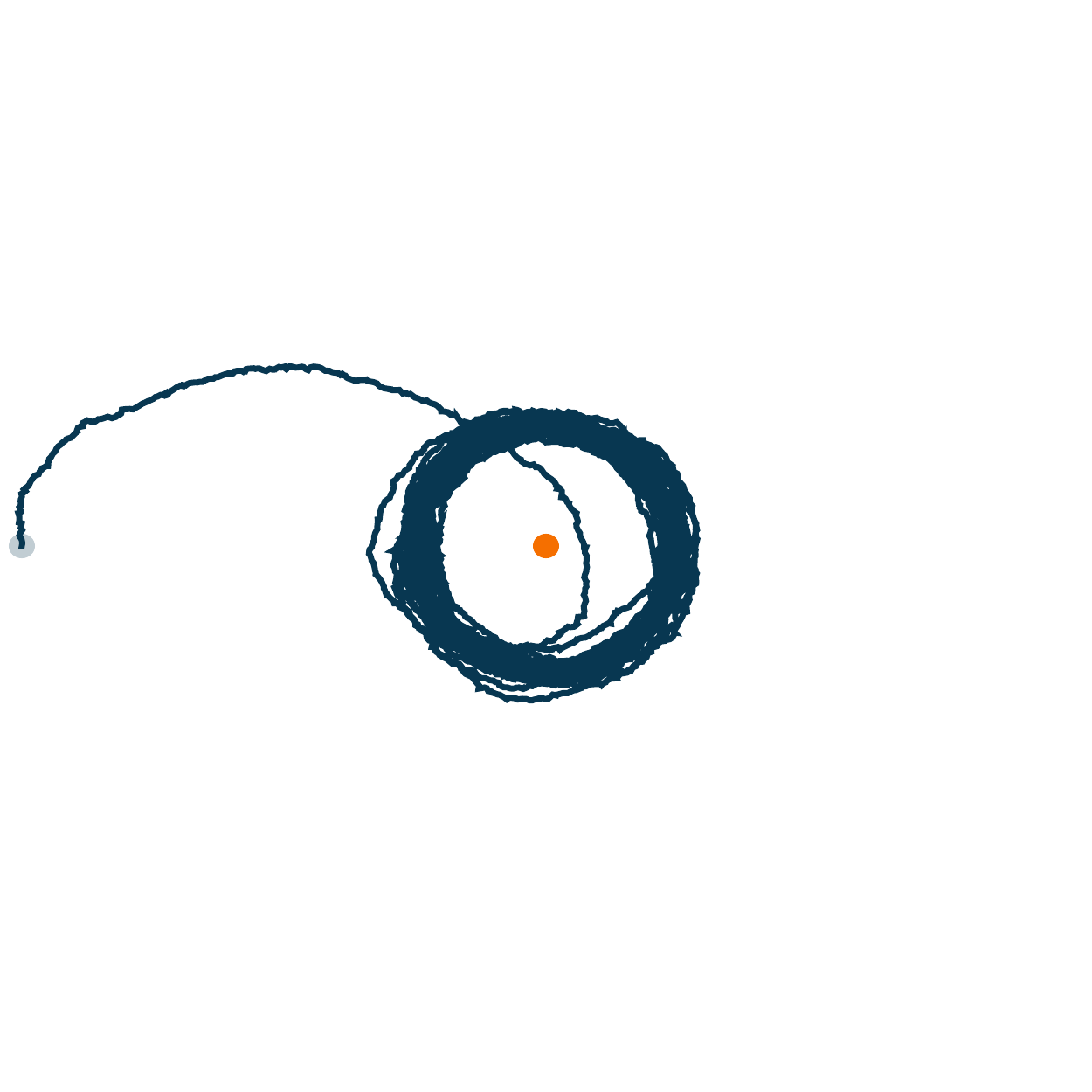}
    \includegraphics[width=0.18\textwidth]{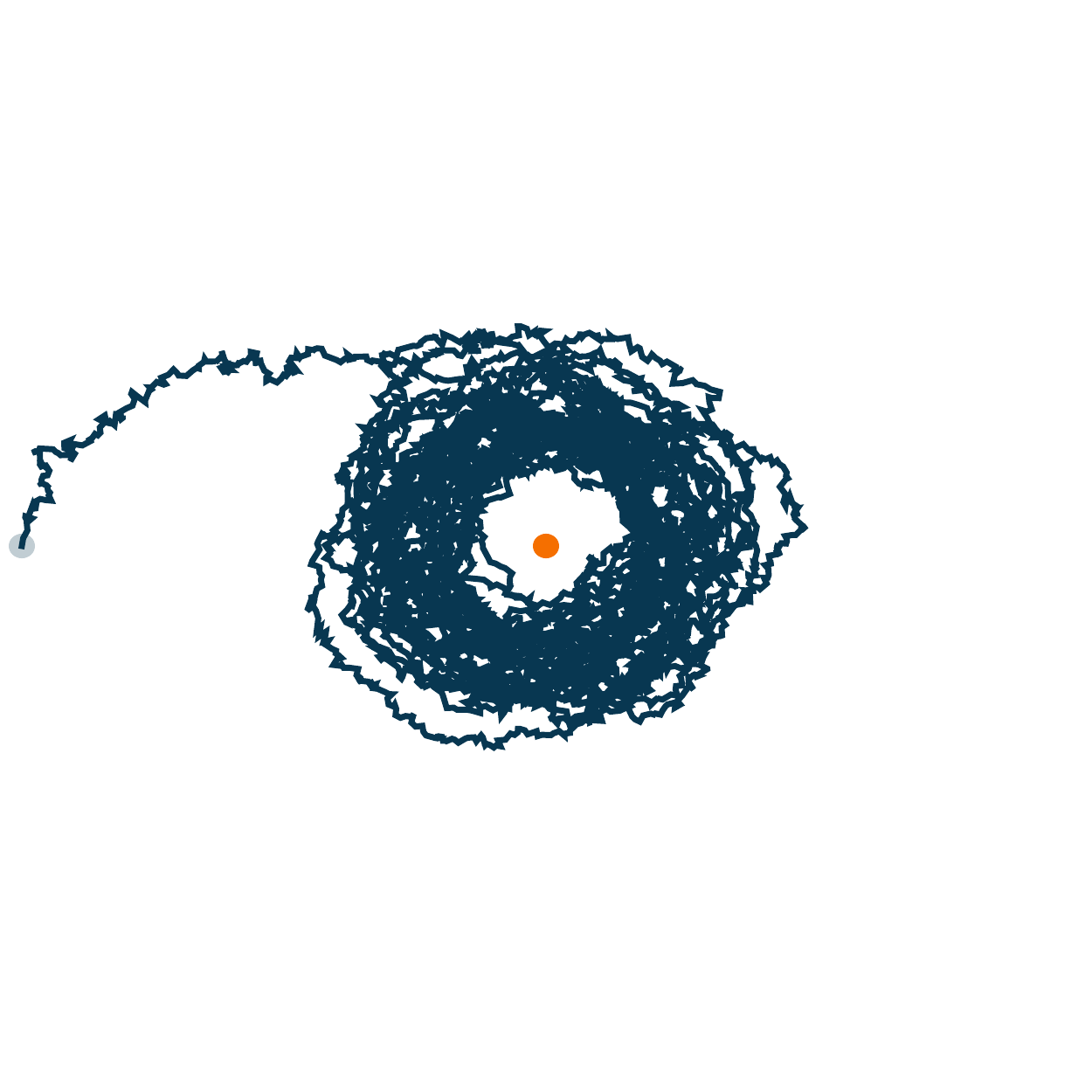}
    \includegraphics[width=0.18\textwidth]{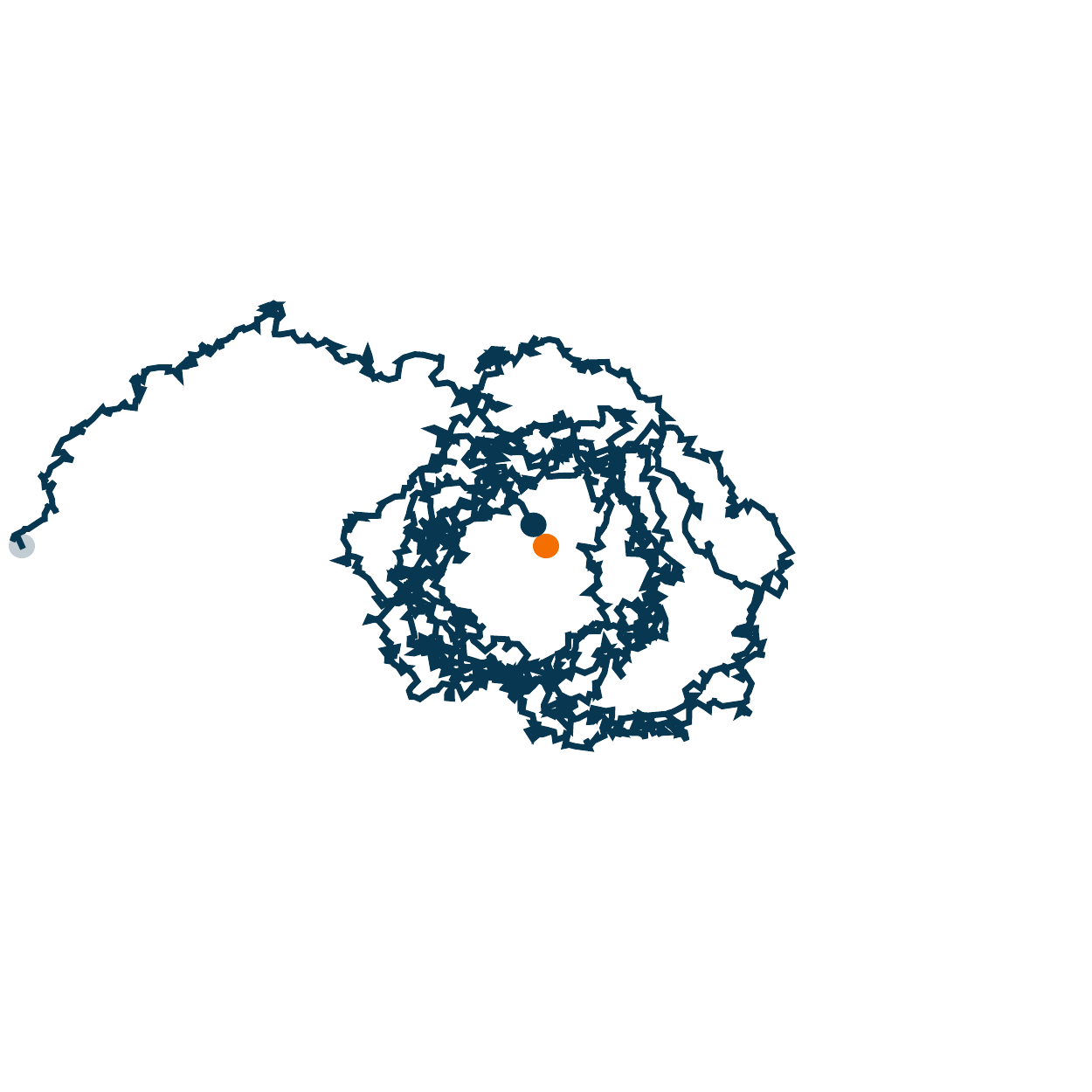}
    \includegraphics[width=0.18\textwidth]{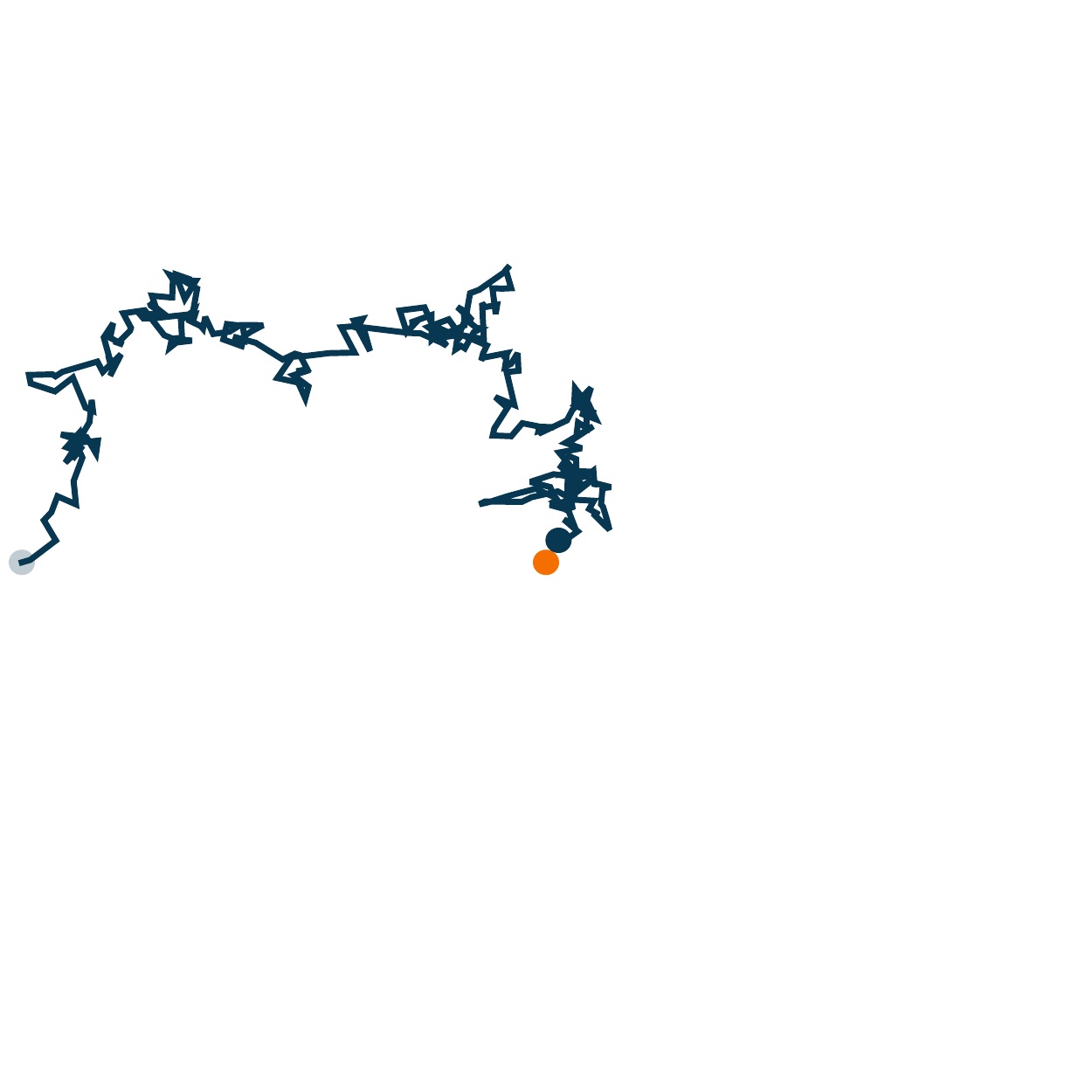} 
    \caption{\textbf{Noisy single-particle trajectories.}
    All trajectories were obtained for $R_0 = 5a$ and starting the particle at $(x_0 = -d,y_0 = 0, \theta_0 = \pi/2)$ (transparent blue disk) with $d = 20a$, and $a$ the diameter of the target (orange disk).
    The dynamics are run either until the particle touches its target, or until time $T_{max} = 1000 a/v_0$.
    The final position of the particle is shown as a dark blue disk.
    Top line: trajectories with rotational noise only, with a noise amplitude that grows from left to right (panels were obtained with $Pe_r = 10000, 1000, 100, 10,$ and $1$).
    Bottom line: similar plots obtained with translational noise only (from left to right, $Pe = 1000, 100, 10, 5,$ and $1$).}
    \label{fig:NoisyParticleTrajectories}
\end{figure}

All in all, this simple approximation suggests that even in the single-particle regime there is a transition, when tuning the amplitude of the noises, between trajectories that still orbit around the target but with a finite width, and trajectories that reach the target following a diffusive path.
This is supported by the shape of single-particle trajectories, like those shown in Fig.~\ref{fig:NoisyParticleTrajectories}: when one switches on a source of noise, the orbits simply get wider at first, but then disappear altogether.
Note that this transition is reminiscent of the trajectories of real-life self-propelled particles in harmonic confinement, that feature a transition between trajectories that spend a long time at the bottom of the potential and others in which particles orbit around the minimum but never visit it~\cite{Dauchot2019}.

\end{document}